\documentclass[11pt,a4paper]{article}
\usepackage{jcappub}
\usepackage{graphicx}
\usepackage{amsmath}
\usepackage{bm}
\usepackage{multirow}
\usepackage{comment}

\newcommand{\simgt}{\lower.5ex\hbox{$\; \buildrel > \over \sim \;$}}
\newcommand{\simlt}{\lower.5ex\hbox{$\; \buildrel < \over \sim \;$}}
\newcommand{\solM}{\mathrm{M_{\odot}}}

\begin{document}

\title{Testing gravity with halo density profiles observed through 
gravitational lensing}

\author{Tatsuya Narikawa and Kazuhiro Yamamoto}
\affiliation{Department of Physical Science, Hiroshima University, 
Higashi-Hiroshima 739-8526,~Japan}
\emailAdd{narikawa@theo.phys.sci.hiroshima-u.ac.jp, kazuhiro@hiroshima-u.ac.jp}

\abstract{
We present a new test of the modified gravity endowed with the Vainshtein
mechanism with the density profile of a galaxy cluster halo observed 
through gravitational lensing.
A scalar degree of freedom in the galileon modified gravity 
is screened by the Vainshtein mechanism to recover Newtonian gravity 
in high-density regions, however it might not be completely hidden 
on the outer side of a cluster of galaxies. 
Then the modified gravity might yield 
an observational signature in a surface mass density of a cluster of 
galaxies measured through gravitational lensing, since the scalar field 
could contribute to the lensing potential. 
We investigate how the transition in the Vainshtein mechanism affects  
the surface mass density observed through gravitational lensing, assuming 
that the density profile of a cluster of galaxies follows the original 
Navarro-Frenk-White (NFW) profile, the generalized NFW profile 
and the Einasto profile. 
We compare the theoretical predictions with observational results
of the surface mass density reported recently by other researchers. 
We obtain constraints on the amplitude and the typical scale of the 
transition in the Vainshtein mechanism in a subclass of the generalized 
galileon model. 
}
\keywords{modified gravity, gravitational lensing}

%\arxivnumvber{1201.4037}

%\begin{document}

\maketitle

\def\bfx{{\bf x}}
\def\mpl2{{M_{\rm Pl}^2}}
\def\hmpc{h^{-1}{\rm Mpc}}
%=========================================================================
%=========================================================================
\section{Introduction}
%=========================================================================
Discovery of the late-time accelerated expansion of our universe 
\cite{Riess,Perlmutter} boosted interests in nature of gravity
on cosmological scales, which might be a key to explore its
origin (e.g., Refs.~\cite{DurrerMaartens,JainKhoury,Tsujikawa10}). 
Modification of gravity is an alternative approach to dark energy 
paradigm to explain the cosmic accelerated expansion 
\cite{DurrerMaartens,JainKhoury,Tsujikawa10,DE06,AmeTsujiDE}.
There have been proposed, for example,  $f(R)$ model 
\cite{HuSawicki,Starobinsky,ApplebyBattye,NojiriOdintsov,fRreview}, 
Dvali-Gabadadze-Porrati (DGP) model~\cite{DGP}, 
and the galileon model~\cite{GALMG}.
In general, it is very challenging to construct a viable model that
explains the late-time accelerated expansion of the background 
universe, satisfying the local gravity constraints without 
theoretical plague simultaneously.
%
%The $f(R)$ model is constructed by replacing the Einstein-Hilbert action with a general function of the Ricci scalar $R$. 
%

The DGP model is described in the context of the braneworld scenario, 
which consists of a 3+1-dimensional brane embedded in a 5-dimensional 
bulk. This model has an interesting phenomenology, which yields two 
branches of the Friedmann equation, a self-accelerating branch (sDGP) 
and a normal branch (nDGP).
In the sDGP branch, the expansion of the background universe 
self-accelerates at late times without a cosmological constant, 
while the nDGP branch needs to add a 
stress-energy component with negative pressure on the brane 
to be consistent with cosmological observations~\cite{DGP2,DGP3}. 
Unfortunately, the sDGP model inevitably gives rise to a ghost 
\cite{Nicolis04,Gorbunov06}.
Moreover, the sDGP model is disfavored by the cosmological 
observations~\cite{Fairbairn,Maartens06,Song07}.

Inspired by the decoupling limit of the DGP model~\cite{Luty03,Nicolis04}, the galileon 
gravity theory has been studied as a possible alternative to 
large distance modification of gravity (e.g.,~\cite{GALMG,CG,GC,SAUGC,ELCP,CEGH,GInflation,DeffayetKGB,GGC,CCGF,OCG,KimuraKGB}).
This theory introduces a scalar field invariant under the 
Galilean shift symmetry 
$\partial_\mu\phi\rightarrow\partial_\mu\phi+b_\mu$ in the Minkowski space-time, 
which keeps the equation of motion being the second order differential 
equation. 
Although the Lagrangian no longer satisfies the Galilean shift symmetry 
in a curved spacetime, one can construct the generalized galileon model
whose field equation remains a second order differential equation, 
which simultaneously admits the self-accelerating solution in a FRW 
universe without a ghost instability~\cite{Horndeski,Deffayet,GGI}.

In general, the gravitational force and the gravitational potential 
are modified  on the scales of cosmology as a result of the 
modification of gravity. 
However, the solar system observations require that Newtonian gravity 
is recovered on those scales~\cite{Will06}. 
There have been proposed a few successful screening mechanisms
depending on modified gravity models, e.g.,  
the chameleon mechanism~\cite{MotaBarrow,Chameleon1,Chameleon2}, 
the symmetron~\cite{symmetron,sc}, 
%
%The chameleon mechanism is that the mass of a scalar degree of freedom 
%is large in dense regions and scalar field is screened.
%
%The symmetron, which is a scalar field associated with the dark sector 
%whose coupling to matter depends on the surrounding matter density, 
%is decoupled and screened within dense regions.
and the Vainshtein mechanism~\cite{Vainshtein}.
One of the notable features of the galileon gravity is the Vainshtein 
mechanism, which hides the scalar field to recover Newtonian gravity
in high-density regions.
In the Vainshtein mechanism, the self-interaction term like
$(\partial\phi)^2\square\phi$ plays an important roll.
Due to the nonlinear interaction, the scalar field is 
suppressed in high-density regions, hence it is screened there.
Thanks to this screening mechanism, the galileon gravity models 
can evade the solar-system constraints~\cite{GALMG} 
(c.f.~\cite{Kimura2011}).

It is worth examining whether the screening mechanism works 
completely or not around high-density regions on cosmological scales.
The structure of halos of galaxy and cluster of galaxies may be useful 
to test these modified gravity models. 
Recently, several researchers have investigated such a possibility of 
testing these modified gravity models on the scales of halos of galaxy
and cluster of galaxies~\cite{Galihalo,dynamicalmass,Wyman,Burikham,Zhao11,LiHu,Clampitt,EPI,SS,GB,dwarfgalaxy,Lombriser2011,HuiNicolice2012}. 
In \cite{sjorsmortsell}, the authors have investigated a 
constraint on the Vainshtein mechanism in the context of the massive gravity,
using gravitational lensing and velocity dispersion data from galaxies.
In a recent work, we investigated how 
the transition in the Vainshtein mechanism may appear in the sDGP model 
and the galileon model~\cite{Galihalo}. 
We found that the circular speed of a test particle in the sDGP and 
the galileon model deviates from that in Newtonian gravity 
at $10\%$ level on the outer region of a halo with the Navarro-Frenk-White 
(NFW) density profile (e.g.,~\cite{NFW1996,NFW1997}).
Thus the Vainshtein mechanism might not completely hide the effect
of the modified gravity on the cluster's scales.

The purpose of the present paper is to investigate how the transition 
in the Vainshtein mechanism in a subclass of the generalized galileon
model appears in an observation of a halo density profile measured 
through the gravitational lensing phenomena.
We consider a subclass of the most general second-order scalar-tensor 
theory~\cite{Deffayet,GGI,DeFelice2011,Horndeski}, in which the scalar 
field modifies the equation connecting the lensing potential and the 
matter density.
In this model, the observational quantities measured through the lensing 
phenomena are contaminated by the scalar field's effect, which 
might leave testable signatures in observed halos.
Then, we focus our investigation on the signature of the modified 
gravity by comparing theoretical predictions and observations.

Many works have been done for testing gravity on cosmological scales 
using, for example, the large scale structures of galaxies and
the redshift-space distortions
(e.g.,~\cite{Shirata05,Shirata07,YNHNS,Reyes,Guzzo,Yamamoto2008}), 
and the integrated Sachs-Wolfe effect through cross-correlations between the 
cosmic microwave background temperature anisotropies and the galaxy 
distributions (e.g.,~\cite{KimuraISW}). 
The cluster abundance (e.g.,~\cite{SchmidtAbundance}) provides a test of 
gravity on rather smaller scales, but it reflects the regime of the 
linear evolution of density perturbations substantially.
On smaller scales, other tests of deviation from general relativity
have been done~\cite{Bolton06,ISL07,TSmith09,Uzan10,Schwab10,Wojtak}. 
A halo of cluster of galaxies provides a unique chance to test
the gravity theory on the scale between the solar-system and 
the large scale structure of galaxies, where the nonlinear effect of 
the density perturbations plays an important role.

%{\bf Gravitational lensing}
In the present paper, we focus our investigation on the 
surface mass density of cluster halo measured with 
gravitational lensing.
Cluster lensing surveys are in progress, which is useful to obtain 
mass distribution of clusters over a wide range of radius by combing 
strong and weak lensing data. 
Recently, measurements of the surface mass density 
were reported using the gravitational lensing 
\cite{Lombriser2011,Umetsu2011a,Umetsu2011b,Oguri2011}. 
The error of the stacked data is small, which is useful to test 
the modified gravity as will be demonstrated below. 
Umetsu et al. derived a mean surface mass profile of four clusters, 
A1689, A1703, A370, and Cl0024+17, from Hubble Space Telescope and 
Subaru images, in the range $R=(40-2800)h^{-1}{\rm kpc}$ at the mean 
lensing redshift $\langle z_l \rangle = 0.32$ 
\cite{Umetsu2011a,Umetsu2011b}.
They obtained mass profiles with high-precision, by combining independent 
strong-lensing measurements, weak-lensing distortion, and magnification.
Oguri et al. presented the mass distribution of a sample of 25 galaxy 
clusters, in the range $R=(63-5010)h^{-1}{\rm kpc}$ at the mean 
lensing redshift $\langle z_l \rangle = 0.47$ and the mean 
source redshift $\langle z_s \rangle =1.1$~\cite{Oguri2011}.
The cluster sample is based on the Sloan Giant Arcs Survey (SGAS) 
from the Sloan Digital Sky Survey (SDSS).
They derived the differential surface mass density by combining the 
strong lensing 
information from the giant arcs and weak lensing measurements 
from Subaru/Suprime-Cam~\cite{Miyazaki} images.
Umetsu et al. and Oguri et al. fit the data within the general 
relativity, and they found that the NFW profile favors the data.

We confront the {\it observed} surface mass density of clusters 
with the theoretical prediction of the modified gravity model
endowed with the Vainshtein mechanism. 
In the theoretical modeling, we introduce the parameters 
$\mu$ and $\epsilon$, respectively, which characterize the amplitude and the 
typical radius of the transition in the Vainshtein mechanism. 
We obtain constraints on $\mu$ and $\epsilon$. 
This demonstrates the usefulness of the 
surface mass density to constrain the modified gravity model.

The structure of this paper is as follows. 
In section \ref{sec:formulation}, we derive our basic theoretical 
formulas in a subclass of the generalized galileon modified gravity 
to compare with observational quantities from gravitational lensing.  
In section \ref{sec:profile}, we scrutinize the theoretical behavior of 
the surface mass density profile depending on familiar density profiles. 
In section \ref{sec:results}, we present the results of the  
confrontation between the theoretical predictions and the observational 
results, which derive constraints on the model parameters $\mu$
and $\epsilon$ characterizing the modification of gravity.
%We also discuss about the results and interpretations are also given there.
In section \ref{sec:discussion}, we discuss about the results and our  
interpretations.
Section \ref{sec:summary} is devoted to summary and conclusions.
In appendix \ref{sec:coefficient}, we summarize the definitions of 
the coefficients in the perturbation equations in section \ref{sec:formulation}.
%In appendix \ref{sec:kinetic}, we discuss theoretical allowed region of $\epsilon$
%for the kinetically driven model.
In appendix \ref{sec:Galileon}, we summarize the coefficients 
in the perturbation equations in the original galileon model.

Throughout this paper, we use units in which the speed of light equals unity,
and we follow the convention $(-,+,+,+)$.
We use the reduced Planck mass $M_{\rm Pl}$, defined by
$M_{\rm Pl}=1/\sqrt{8\pi G}$ with Newton's gravitational constant $G$.
We adopt the Hubble constant $H_0=100~h{\rm km/s/Mpc}$ with $h=0.702$, and
the matter density parameter at present $\Omega_0=0.275$~\cite{WMAP7}.
% and $\rho_{cr,0}=2.775\times10^{11}/h[\solM(h{\rm Mpc}^{-1})^3]$.

%=========================================================================
%=========================================================================
\section{Formulation}\label{sec:formulation}
%=========================================================================
%=========================================================================
%=========================================================================
\subsection{Modified gravity model}\label{sec:MG}
%=========================================================================
%=========================================================================

%=========================================================================
%=========================================================================
%\subsubsection{Second-order scalar-tensor theories}
%=========================================================================
The most general second-order scalar-tensor theory was derived 
by Horndeski~\cite{Horndeski} for the first time, which was 
recently rediscovered by Deffayet et al.~\cite{Deffayet} as the 
most generalized galileon theory, which contains four arbitrary
functions of the functions of $\phi$ and 
$X=-g^{\mu\nu}\nabla_\mu\phi\nabla_\nu\phi/2$. 
The Vainshtein mechanism in the generalized galileon theory is
discussed recently~\cite{Kimura2011}, which clarified variety 
of the solutions of the scalar field in the generalized theory.
We here consider a subclass of the general second-order 
scalar-tensor theories in a curved spacetime, which is 
nonminimally coupled to gravity with the action~\cite{DeFelice2011},
\begin{eqnarray}
S=\int d^4x\sqrt{-g}\left[{{1\over 2}F(\phi)R}+K(\phi,X)
-G(\phi,X)\square\phi+{\cal L}_{\rm m}\right],
\label{action}
\end{eqnarray}
where $K(\phi,X)$ and $G(\phi,X)$ are the arbitrary functions 
of $\phi$ and $X$, $F(\phi)$ is the function of $\phi$, and 
${\cal L}_{\rm m}$ is the matter Lagrangian. 
We assume that the matter fields do not have direct couplings 
with the field $\phi$.
This corresponds to the Lagrangian with 
$G_4=F(\phi)/2$ and $G_5=0$ in the most general galileon model 
\cite{GGI}.
{}The action reduces to the kinetic gravity braiding model, 
for the choice $F(\phi)=\mpl2$~\cite{DeffayetKGB,KimuraKGB}.
{}The original galileon model is reproduced by choosing 
$F(\phi)=\mpl2$, $K(\phi,X)=-X$ and $G(\phi,X)={(r_c^2/{M_{\rm Pl}})}X$, 
where $r_c$ is the parameter~\cite{DeffayetKGB}.

%=========================================================================
%\subsection{Perturbation equations in the quasi-static approximation}\label{sec:be}
%=========================================================================
In this subsection, we summarize the perturbation equations for 
gravity and the scalar field in the cosmological background. 
In reference \cite{Kimura2011}, the Vainshtein mechanism 
in the most general second-order scalar-tensor theory was investigated. 
The model in the present paper (\ref{action}), a subclass of
the most general second-order scalar-tensor theory, was also investigated
therein. Following \cite{Kimura2011},
we choose the Newtonian gauge, 
\begin{eqnarray}
%  ds^2&=&-(1+2\Psi(t,\bfx))dt^2+a(t)^2(1+2\Phi(t,\bfx))
%  \bigl(dx^2+dy^2+dz^2\bigr)\\
ds^2  &=&-(1+2\Psi(t,\bfx))dt^2+a(t)^2(1+2\Phi(t,\bfx))\delta_{ij}dx^i dx^j,
\label{Ngauge}
\end{eqnarray}
where $a(t)$ is the scale factor.
Within the subhorizon scales with the quasi-static approximation,  
we have the following perturbed equations,
\begin{eqnarray}
&&{\triangle\over a^2}\Phi=
-4\pi G \delta\rho +\xi{\triangle\over a^2}\varphi \label{pE}, 
\label{pE1}
\\
&&\Phi+\Psi=-\alpha\varphi,
\label{pE2}
\end{eqnarray}
and 
\begin{eqnarray}
{\triangle\over a^2}\varphi+\lambda^2\left({\varphi_{,ij}\over a^2}{\varphi^{,ij}\over a^2}- \left({\triangle\over a^2}\varphi\right)^2\right)=-4\pi G \zeta \delta\rho,
\label{gE}
\end{eqnarray}
where $\varphi(\bfx)$ denotes the perturbation of scalar field defined by 
$\phi(t,\bfx)=\phi(t)(1+\varphi(\bfx))$,
$\triangle$ represents the Laplace differentiation operator,
$\delta\rho$ is the perturbed matter density,
the Newton's gravitational constant is defined by 
$G \equiv 1/(8\pi F(\phi))$, 
and the coefficients $\alpha$, $\xi$, $\zeta$, and $\lambda^2$ 
depend on the functions $F(\phi)$, $K(\phi,X)$, and $G(\phi,X)$, whose
explicit expressions are summarized in appendix \ref{sec:coefficient}.

The above equations are derived as follows (see \cite{Kimura2011} for details).
We consider the gravitational and scalar fields on subhorizon
scales sourced by a non-relativistic matter density perturbation $\delta\rho$. 
Then, we may ignore time derivative in the field equations, while
keeping spatial derivatives. 
Denoting the quantities for perturbations $\Psi$, $\Phi$ and $\varphi$
by ${\cal V}$, we keep all the terms $(\partial^2 {\cal V})^n$
with $n\geq1$, because the $L^2(t)\partial^2{\cal V}$ could be 
large on small scales, where $L(t)$ is a typical length scale
associated with the background evolution which may be as large
as the Hubble radius. 
Within this framework, it has been shown that the terms 
$(\partial^2 {\cal V})^n$ with  $n\geq2$ do not appear 
in the gravitational field equations (\ref{pE1}) and (\ref{pE2}) 
for our model (\ref{action}). 
But only the terms of $(\partial^2 {\cal V})^2$ appear in the scalar 
field equation (\ref{gE}). 

Note that the coefficients $\alpha$, $\xi$, $\zeta$, and 
$\lambda^2$ are determined by the background field evolution 
depending on $F(\phi)$, $K(\phi,X)$, and $G(\phi,X)$. 
The coefficients may take any value because $F(\phi)$,
$K(\phi,X)$, and $G(\phi,X)$ are {\it arbitrary} functions of 
$\phi$ and $X$. We should choose $F(\phi)$,  $K(\phi,X)$, and 
$G(\phi,X)$  requiring the accelerated expansion of the 
universe and satisfying the conditions of no ghost nor instability, 
which will limit the arbitrary functions. 
However, we here assume that $\alpha$, $\xi$, $\zeta$, and 
$\lambda^2$ may take any value reflecting the arbitrariness
of the functions in the model, and consider the constraint
from observation through gravitational lensing.

Combining (\ref{pE1}) and (\ref{pE2}), we have
\begin{eqnarray}
&&{\triangle\over a^2}\left({\Psi-\Phi \over2}\right)=4\pi G \delta\rho-{\alpha+2\xi\over2}{\triangle\over a^2}\varphi.
\label{eqb}
\end{eqnarray}
In the spherically symmetric case, (\ref{gE}) yields
\begin{eqnarray}
{d\varphi\over dr}= {r \over 4\lambda^{2}}
\left(1-\sqrt{1+{8G\lambda^{2}\zeta M(r)\over r^{3}}}\right),
\label{eq:G_Field}
\end{eqnarray}
where we defined the enclosed mass
$M(r)=4\pi\int_0^rdr'r'^2\delta\rho(r')$.
In the spherically symmetric case, 
(\ref{eq:G_Field}) leads to
\begin{eqnarray}
{\triangle\over a^2}\varphi={1\over r^2}{d\over dr}\left(r^2{d\varphi\over dr}\right)=
{3\over4\lambda^2}\left(1-\sqrt{1+{8 G \lambda^2\zeta M(r)\over r^3}}\right)-G\zeta {4\pi r^3\delta\rho(r)-3M(r)\over r^3\sqrt{1+{8G \lambda^2\zeta M(r)\over r^3}}}.
\label{eq:trianglephi}
\end{eqnarray}
%where we choose the solution with the boundary condition that $\varphi$ vanishes at $r\rightarrow \infty$.
%
%\begin{eqnarray}
%\triangle \Psi-\triangle \Phi =8\pi G\rho 
%-(\alpha+2\xi)\triangle \varphi
%\end{eqnarray}
%
%Note that the lensing potential in DGP model is not modified because of $\alpha+2\xi=0$.

As we will show in the below (see also \cite{Kimura2011}), 
the lensing signature under the influence of the Vainshtein mechanism
in our model (\ref{action}) is described by the two combination 
of the parameters $\lambda^2\zeta$ and $(\alpha+2\xi)\zeta/2$, 
on which we focus our investigation. 
In the original galileon model, we have non-zero values of 
$\lambda^2\zeta$ and $(\alpha+2\xi)\zeta/2$ (see appendix B). 
On the other hand, in the model of the sDGP model, we have $\alpha=-1$ 
and $\xi=1/2$, then $(\alpha+2\xi)\zeta/2$ reduces to zero, 
while $\lambda^2\zeta$ takes a non-zero value.\footnote{
In the spatially flat sDGP model, we have 
$\lambda^2\zeta=2(r_c/3\beta)^2$, where $r_c=1/[(1-\Omega_m)H_0]$ 
and $\beta=1-2Hr_c(1+\dot H/3H^2)$.}
Therefore, in the sDGP model, the relation between the lensing potential 
$(\Phi-\Psi)/2$ and the matter density perturbation (\ref{eqb}) 
is the same as that in the general relativity.
In this case, our method cannot put a constraint because 
there is no effect of the modified gravity on the gravitational 
lensing in such a class of models. 

In order to obtain the closed system of equations, we need
another equation for the matter density perturbation $\delta\rho$.
Because we are considering the model in which the matter 
is minimally coupled with the scalar field, the matter 
component follows the usual equation under the influence of 
the gravitational potential $\Psi$. 
Instead of specifying the coupled equations of $\delta\rho$ and $\Psi$, 
we simply assume the NFW density profile or other well-known 
profiles for the matter density perturbation $\delta\rho$ 
(cf. \cite{Galihalo}).  
From N-body simulations in reference~\cite{dynamicalmass}, 
it is suggested that the NFW profile describes the density 
profile of halos in the DGP model though the parameters of 
the profiles are scaled. This is a supporting evidence of 
our assumption in the present paper.

%=========================================================================
%=========================================================================
\subsection{Gravitational lensing in modified gravity}
%=========================================================================
Now let us consider observational quantities observed through
gravitational lensing. The light propagation in the 
Newtonian gauge (\ref{Ngauge}) is described by the 
lensing potential $(\Phi-\Psi)/2$.
For example, the convergence $\kappa(\chi)$ of the gravitational 
lensing is given by (e.g.,~\cite{Bartelmann2001,Dodelson}) 
\begin{eqnarray}
%\gamma_1 &=& -{1\over 2} \int_0^\chi d\chi' 
% {f_K(\chi-\chi')\over f_K(\chi)}f_K(\chi')
%\left[(\Phi-\Psi),_{11}-(\Phi-\Psi),_{22}\right],
%\\
%\gamma_2 &=& - \int_0^\chi d\chi' 
% {f_K(\chi-\chi')\over f_K(\chi)}f_K(\chi')
%\left[(\Phi-\Psi),_{12}\right],
%\\
%{\rm and}
%\\
\kappa &\simeq& -{1\over 2} \int_0^\chi d\chi' 
 {f_K(\chi-\chi')f_K(\chi')\over f_K(\chi)}
\triangle^{(2D)}(\Phi-\Psi),
\label{Ekappa}
\end{eqnarray}
where $f_K(\chi)$ is the comoving angular diameter distance and 
$\triangle^{(2D)}$ is the comoving two dimensional Laplacian.
Note that $f_K(\chi) = \chi$ for the spatially flat universe.
The shear can be written in a similar form with (\ref{Ekappa}).
It is worthy to note that any quantity of gravitational lensing 
is described by the lensing potential.

Using the thin lens approximation, (\ref{Ekappa}) and (\ref{eqb}) 
yield
\begin{eqnarray}
\kappa &\simeq&
{f_K(\chi_{\rm S}-\chi_{\rm L})f_K(\chi_{\rm L})\over f_K(\chi_{\rm S})}
\int_0^{\chi_{\rm S}} d\chi' \left[4\pi G\rho(r') -{\alpha+2\xi \over 2}
{\triangle\over a_{\rm L}^2} \varphi \right]a_{\rm L}^2,
\end{eqnarray}
where $\chi_{\rm S}$ is the comoving distance between the 
observer and the source, $\chi_{\rm L}$ is the comoving 
radial distance between the observer and the lens object, 
and $a_{\rm L}$ is the scale factor at which we observe the lens object. 
%
%You know
%\begin{eqnarray}
%\kappa = {\Sigma_{\rm S}(R)\over\Sigma_{\rm crit}},
%\end{eqnarray}
We define the {\it observed} surface mass density, which 
is obtained through the gravitational lensing phenomena, 
%\begin{eqnarray}
%\Sigma_{\rm S}(R)={1\over8\pi G}
%\int_0^{\chi_{\rm S}} d\chi' 
%\left[(\Psi-\Phi),_{11}+(\Psi-\Phi),_{22}\right].
%\end{eqnarray}
%Using the thin lens approximation ($(\Psi-\Phi),_{33}=0$),
%\begin{eqnarray}
%{1\over 2} \int_0^{\chi_{\rm S}} d\chi' 
%\left[(\Psi-\Phi),_{11}+(\Psi-\Phi),_{22}\right]
%&\simeq&
%{1\over 2} \int_0^{\chi_{\rm S}} d\chi'[\triangle\Psi-\triangle\Phi]
%\end{eqnarray}
%From eq. (\ref{eqb}),
\begin{eqnarray}
\Sigma_{\rm S} = \Sigma_{\rm crit} \kappa
= \int_0^{\chi_{\rm S}} d(a_{\rm L}\chi') \left[\rho(r') -{\alpha+2\xi \over 8\pi G}
{\triangle\over a_{\rm L}^2} \varphi \right],
\label{sigmaS}
\end{eqnarray}
%\begin{eqnarray*}
%\int_0^{\chi_{\rm S}} d\chi'[\triangle\Psi-\triangle\Phi]
%&=&
%\int_0^{\chi_{\rm S}} d\chi' [8\pi G\rho -(\alpha+2\xi)
%\triangle \varphi]=8\pi G \Sigma_{\rm S}
%\end{eqnarray*}
%\begin{eqnarray*}
%\Sigma_{\rm S}=
%\int_0^{\chi_{\rm S}} d\chi' \left[\rho -{3(\alpha+2\xi)
%\over 32\pi G\lambda^2}
%\left(1-\sqrt{1+{32\pi G \rho\lambda^2\zeta \over 3}}
%\right)\right]
%\end{eqnarray*}
where $\Sigma_{\rm crit}$ is defined by 
\begin{eqnarray}
\Sigma_{\rm crit}={1\over4\pi G}{f_K(\chi_{\rm S})\over
f_K(\chi_{\rm L})f_K(\chi_{\rm S}-\chi_{\rm L})a_{\rm L}}.
\label{sigmacrit}
\end{eqnarray}

We introduce the spatial coordinate being the physical coordinate, 
whose origin is located at the center of the lens object by
$Z=a_{\rm L}(\chi-\chi_{\rm L})$ and $r_\perp=a_{\rm L}f_K(\chi_{\rm L})\theta$, 
where $\theta$ is the polar angle of the polar axis connecting the 
observer and the lens object. Then, using (\ref{eq:trianglephi}), 
the {\it observed} surface mass density (\ref{sigmaS}) is written as
\begin{eqnarray}
\Sigma_{\rm S}(r_\perp)
&=&2\int_0^\infty dZ \biggl[\rho(r) -{3(\alpha+2\xi)
\over 32\pi G\lambda^2}
\left(1-\sqrt{1+{8 G \lambda^2\zeta M(r)\over r^3}}\right)
\nonumber \\
&&\hspace{4cm}
+{(\alpha+2\xi)\zeta\over8\pi}{4\pi r^3\rho(r)-3M(r)\over r^3
\sqrt{1+{8G \lambda^2\zeta M(r)\over r^3}}}
\biggr], 
\label{defsigmaS}
\end{eqnarray}
where $r=\sqrt{r_\perp^2+Z^2}$.

%=========================================================================
%=========================================================================
\subsection{Parametrization of the modified gravity}
%=========================================================================

To investigate the constraint on the modified gravity model on the 
halo scales, we introduce the parameters 
$\mu$ and $\epsilon$ as follows, which characterize 
the modification of gravity, 
\begin{eqnarray}
\mu={(\alpha+2\xi)\zeta \over 2},
~~~~~
\epsilon=\sqrt{H_0^2\lambda^2\zeta}.
\end{eqnarray}
%phenomenologically parametrize the modified part by an additional scalar degree of freedom for the surface mass density $\Sigma_{\rm S}(r_\perp)$.
%If it may be inconsistent with gravitational tests on small scales and large scales, we neglect the effects of it to another scales now.
With these two parameters the surface mass density 
(\ref{defsigmaS}) is written as
\begin{eqnarray}
\Sigma_{\rm S}(r_\perp)
=2\int_0^\infty dZ \left[\rho(r) -{3H_0^2\mu \over 16\pi G \epsilon^2}
\left(1-\sqrt{1+{8 G \epsilon^2M(r)\over H_0^2r^3}}\right)
+{\mu\over4\pi}{4\pi r^3\rho(r)-3M(r)\over r^3\sqrt{1+{8G \epsilon^2M(r)\over H_0^2r^3}}}
\right].
\nonumber\\ 
\end{eqnarray}
The physical meaning of $\mu$ and $\epsilon$ is understood as follows.
In the limit of large $r$, (\ref{eq:trianglephi}) and (\ref{eqb}) 
reduce to
\begin{eqnarray}
{\triangle\over a^2}\varphi\simeq-{3G\zeta M(r)\over r^3},
\end{eqnarray}
and 
\begin{eqnarray}
{\triangle\over a^2}\left({\Psi-\Phi \over2}\right)\simeq4\pi G_{\rm eff}\delta\rho,
\end{eqnarray}
respectively, where $G_{\rm eff}=G(1+\mu)$ is the effective gravitational 
constant in the linearized limit. 
%In this limit, we have
%\begin{eqnarray}
%\Sigma_{\rm S}(r_\perp\rightarrow\infty) \simeq 2(1+\mu)\int_0^\infty dZ \rho(r).
%\end{eqnarray}
Thus $\mu$ is the amplitude of the modification of gravity in the outer 
region of a halo in the linearized regime. 
On the other hand, in the limit of small $r$, 
(\ref{eq:G_Field}) gives
\begin{eqnarray}
 \biggl|{d\varphi\over dr}\biggr| \simeq
  {1\over 4\lambda^2} \sqrt{8G\lambda^2\zeta M(r)\over r}\ll
  {G M(r)\over r^2}\sim \biggl|{d\Psi\over dr}\biggr|, ~\biggl|{d\Phi\over dr}\biggr|.
\end{eqnarray}
This means that the effect of the scalar field is screened due to 
the nonlinear interaction of the scalar field in the limit of small 
$r$, which is the Vainshtein mechanism.
Hence, the basic equations (\ref{pE1}), (\ref{pE2}) and (\ref{eqb}) reduce to 
those of general relativity when the scalar field is neglected.

Thus, the modified gravity effect appears in the linear regime at 
large radii, while the modified gravity effect is screened due to the 
nonlinear effect at small radii. 
The transition radius between the two regime is 
the Vainshtein radius $r_V$, 
which we defined by (cf.~\cite{Galihalo,KimuraKGB})
\begin{eqnarray}
 r_V \equiv \left[8G\lambda^2\zeta M_{\rm vir}\right]^{1/3}=
\left[{8G\epsilon^2M_{\rm vir}\over H_0^2}\right]^{1/3}, 
\label{eq:Vainshtein}
\end{eqnarray}
where $M_{\rm vir}$ is the virial mass of a cluster (see below). 
Therefore, $\epsilon$ is the parameter of the Vainshtein radius $r_V$.
In summary, Newtonian gravity is recovered due to the Vainshtein mechanism
for $r\ll r_V$, while the gravity is modified for $r\gg r_V$. 
The Vainshtein radius is rewritten as
\begin{eqnarray}
 r_V = 13.4 \epsilon^{2/3} \left(M_{\rm vir}\over 10^{15}\solM\right)^{1/3} \hmpc.
\end{eqnarray}
In the limit $\mu\rightarrow0$ or $\epsilon\rightarrow\infty$,
Newtonian gravity is reproduced on all scales. 
The original galileon model corresponds to
$\mu=0.26$ and $\epsilon=0.53$ 
$(\mu=0.19$ and $\epsilon=0.43)$ 
at the redshift $0.32$ $(0.47)$, respectively, 
which is the mean redshift of clusters of the observational 
data used below. 

\begin{figure}[bthp]
  \begin{tabular}{cc}
    \begin{minipage}{0.5\textwidth}
      \begin{center}
\includegraphics[width=6.2cm,height=6.2cm,clip]{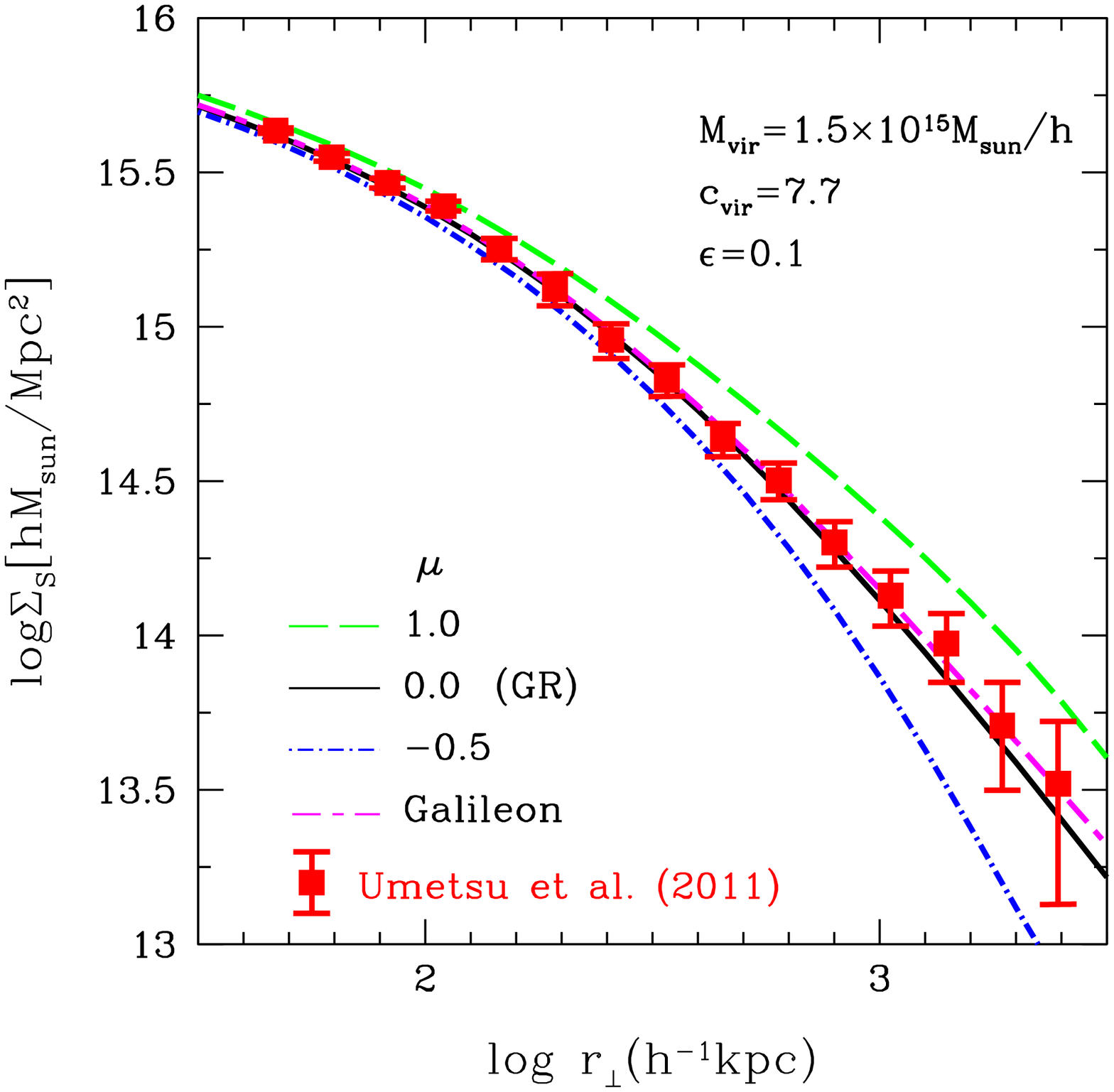}
      \end{center}
    \end{minipage}
    \begin{minipage}{0.5\textwidth}
      \begin{center}
\includegraphics[width=6.2cm,height=6.2cm,clip]{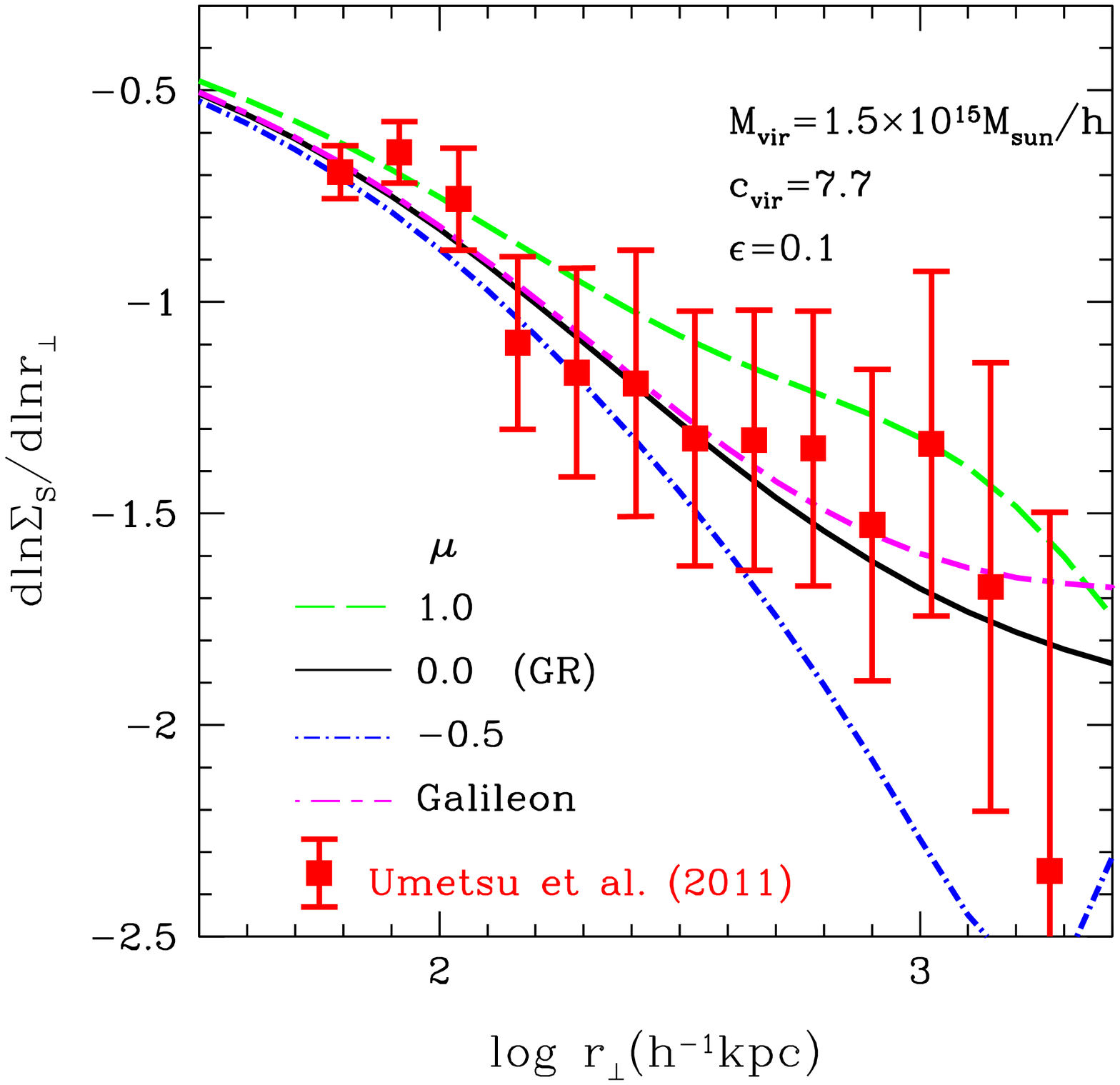}
      \end{center}
    \end{minipage}
  \end{tabular}
\caption{
Surface mass density $\Sigma_{\rm S}(r_\perp)$ (left panel) and the
logarithmic slope $d\ln\Sigma_{\rm S}/d\ln r_\perp$ (right panel) 
as function of $r_\perp$.
The data with the error bar is from Umetsu et al.~\cite{Umetsu2011a,Umetsu2011b}, 
while the curves are the theoretical modeling. 
The (magenta) dot and long-dash curve is the original galileon model 
($\mu=0.26,~\epsilon=0.53$), but the other
curves assume the same value of $\epsilon=0.1$ but the different values 
of $\mu=1$ (green dashed curve), $0$ (black solid curve), and $-0.5$ 
(blue dot and short-dashed curve), respectively. 
Note that $\mu=0$ is Newtonian gravity. 
We here adopted the NFW profile with fixing $M_{\rm vir}=1.5\times10^{15}\solM/h$ and 
$c_{\rm vir}=7.7$. 
}
\label{fig:Sigma.mu.demo}
\end{figure}

\begin{figure}[bhtp]
  \begin{tabular}{cc}
    \begin{minipage}{0.5\textwidth}
      \begin{center}
\includegraphics[width=6.2cm,height=6.2cm,clip]{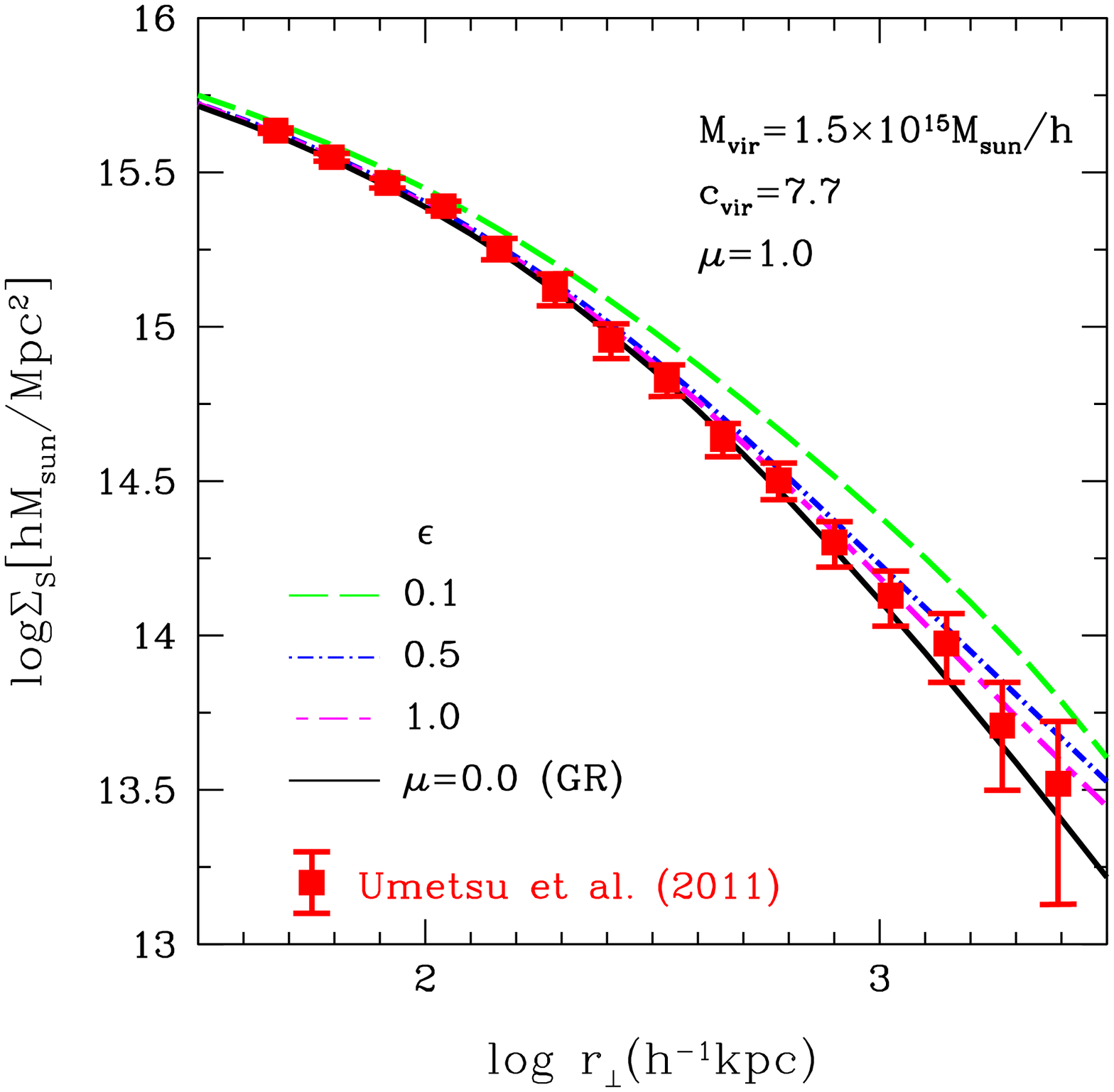}
      \end{center}
    \end{minipage}
    \begin{minipage}{0.5\textwidth}
      \begin{center}
\includegraphics[width=6.2cm,height=6.2cm,clip]{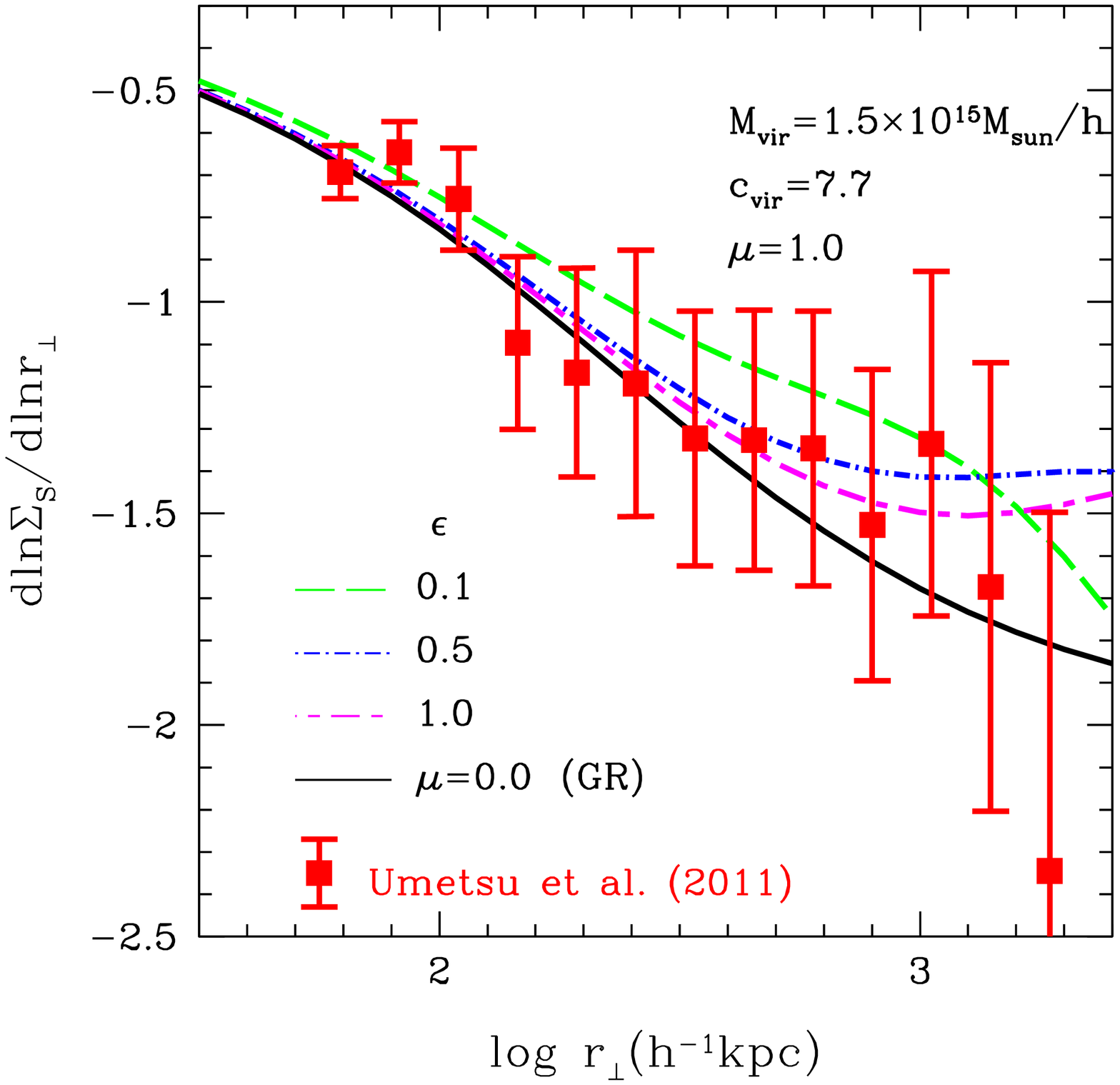}
      \end{center}
    \end{minipage}
  \end{tabular}
\caption{
Same figure 
as figure \ref{fig:Sigma.mu.demo} but with different theoretical models. 
The black solid curve is Newtonian gravity $\mu=0$, the other curves
assume the same value of $\mu=1.0$ and the different values of 
$\epsilon=0.1$ (green dashed curve), $0.5$ (blue dot short-dashed), 
and $1.0$ (magenta dot long-dashed), respectively.
Note that the curves approach Newtonian gravity as $\epsilon$ becomes large. 
We adopted the same NFW profile as figure \ref{fig:Sigma.mu.demo}, 
$M_{\rm vir}=1.5\times10^{15}\solM/h$ and $c_{\rm vir}=7.7$.
}
\label{fig:Sigma.epsilon.demo}
\end{figure}

Figures \ref{fig:Sigma.mu.demo} and \ref{fig:Sigma.epsilon.demo} 
demonstrate typical behavior of the surface mass density 
(left panel) and the logarithmic slope of the 
surface mass density (right panel) as a function of $r_\perp$. 
The curves are the theoretical model parametrized by $\mu$ and 
$\epsilon$, while the points with the error bars show the observational 
data by Umetsu et al. in~\cite{Umetsu2011a,Umetsu2011b}, which is 
obtained by averaging over four massive clusters. 
In figure \ref{fig:Sigma.mu.demo}, the theoretical curves assume
the different values of $\mu$ with fixing $\epsilon$ and the other
parameters for the halo density profile, as described in the
panels, where we assumed the NFW density profile (see the next section). 
This figure demonstrates how the theoretical curve depends on $\mu$.
As $r_\perp$ becomes large, 
the amplitude of $\Sigma_{\rm S}(r_\perp)$ is enhanced for positive 
$\mu$, while it is suppressed for negative $\mu$. 
The (black) solid curve is Newtonian gravity,
while the (magenta) dot and long-dashed curve is the original galileon model. 
Figure \ref{fig:Sigma.epsilon.demo} is the same as figure 
\ref{fig:Sigma.mu.demo} but with adopting the different theoretical 
values. Here the different values of $\epsilon$ are assumed
while $\mu$ and the other parameters for the density profile 
are fixed, as described in this figure. 
As $\epsilon$ describes the Vainshtein radius $r_V$, within which 
Newtonian gravity is recovered, 
$\Sigma_{\rm S}(r_\perp)$ deviates from Newtonian gravity even at small 
radii for small $\epsilon$ (small $r_V$),  
while $\Sigma_{\rm S}(r_\perp)$ deviates from Newtonian gravity only 
at large radii for large $\epsilon$ (large $r_V$).
For the case $\mu<0$, $\Sigma_{\rm S}(r_\perp)$ deviates
from the line of $\mu=0$ in the opposite side of the case $\mu>0$.

%$G_{\rm eff}/G$ is always larger than unity in the galileon model, 
%while there are models with $G_{\rm eff}/G$ that is less than unity.
%Therefore, we investigate the region of $-5.0<\mu<5.0$.
%=========================================================================
%=========================================================================

%epsilon (rX) demo 
%=========================================================================

%Mvir demo 
%=========================================================================
\begin{figure}[bthp]
  \begin{tabular}{cc}
    \begin{minipage}{0.5\textwidth}
      \begin{center}
\includegraphics[width=6.2cm,height=6.2cm,clip]{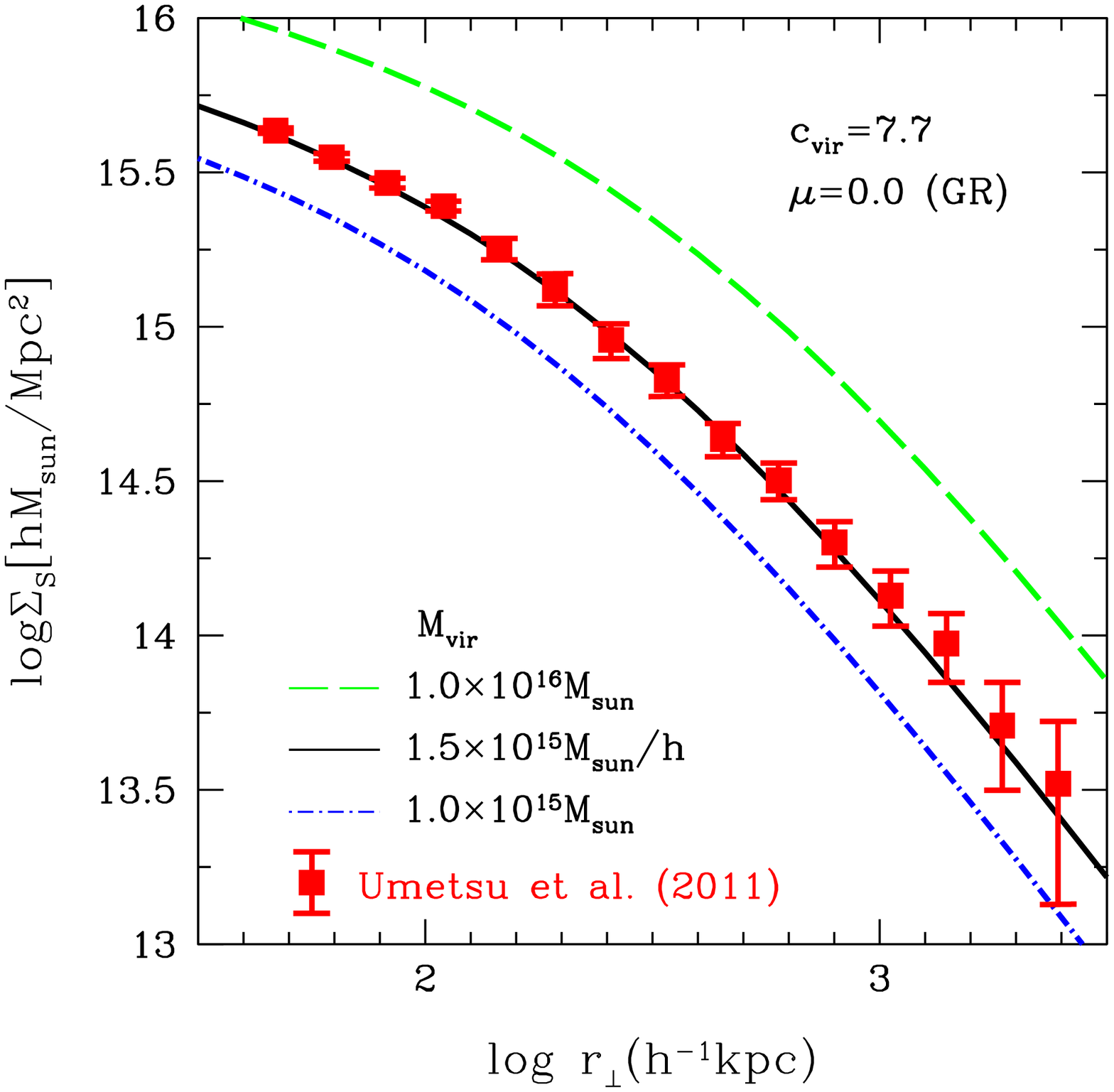}
      \end{center}
    \end{minipage}
    \begin{minipage}{0.5\textwidth}
      \begin{center}
\includegraphics[width=6.2cm,height=6.2cm,clip]{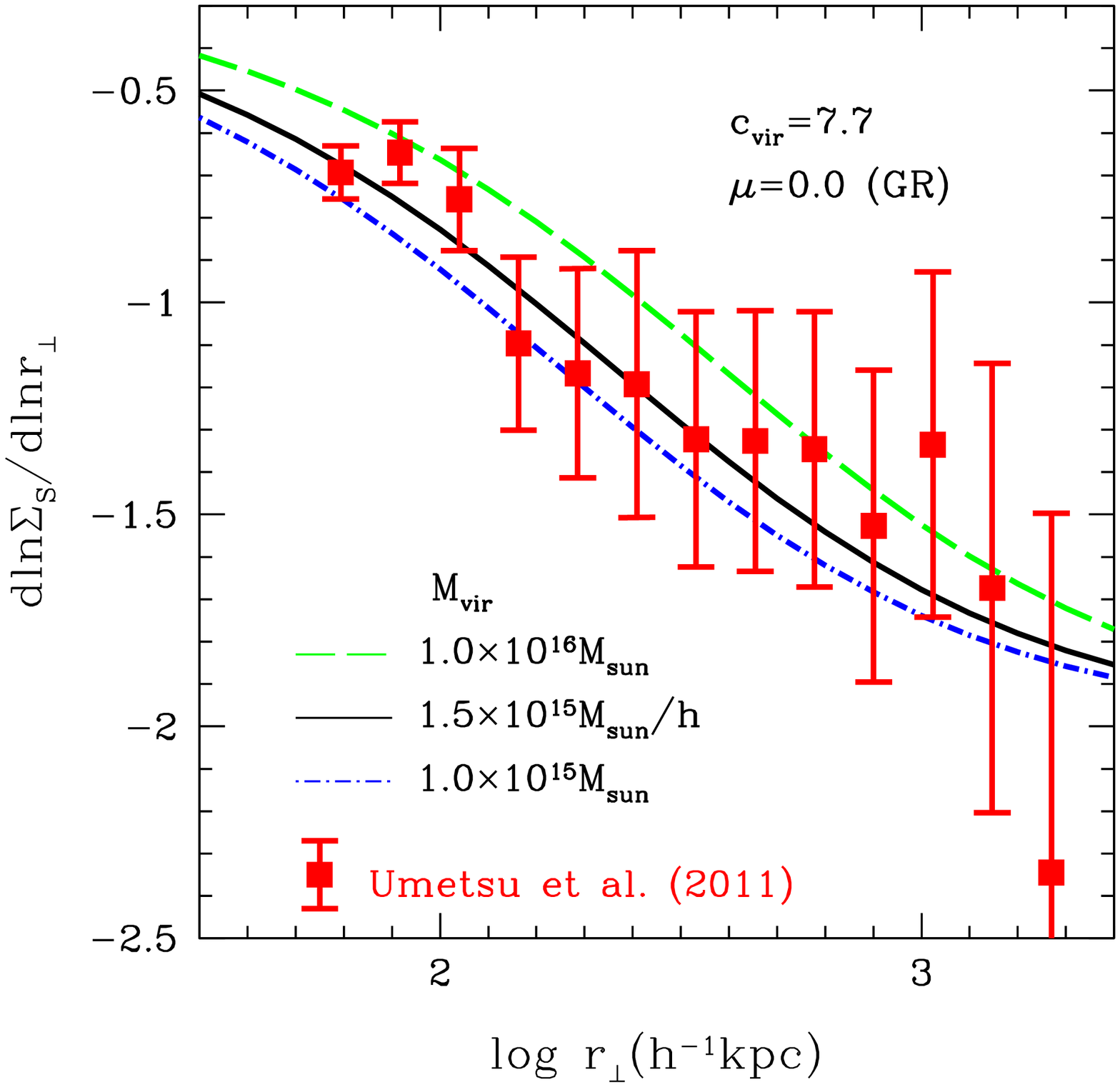}
      \end{center}
    \end{minipage}
  \end{tabular}
\caption{
Surface mass density $\Sigma_{\rm S}(r_\perp)$ (left panel) and the
logarithmic slope $d\ln\Sigma_{\rm S}/d\ln r_\perp$ (right panel) 
as function of $r_\perp$.
The data with the error bar is from Umetsu et al.~\cite{Umetsu2011a,Umetsu2011b}, 
while the curves are the theoretical models assuming Newtonian gravity 
$\mu=0$ and the NFW profile with $c_{\rm vir}=7.7$ and 
$M_{\rm vir}=1.0\times10^{16}\solM$ (green dashed),
$1.5\times10^{15}\solM/h$ (black solid),
and $1.0\times10^{15}\solM$ (blue dot-dashed), respectively. 
}
\label{fig:Sigma.Mvir.demo}
\end{figure}

%cvir demo 
%=========================================================================
\begin{figure}[bhtp]
  \begin{tabular}{cc}
    \begin{minipage}{0.5\textwidth}
      \begin{center}
\includegraphics[width=6.2cm,height=6.2cm,clip]{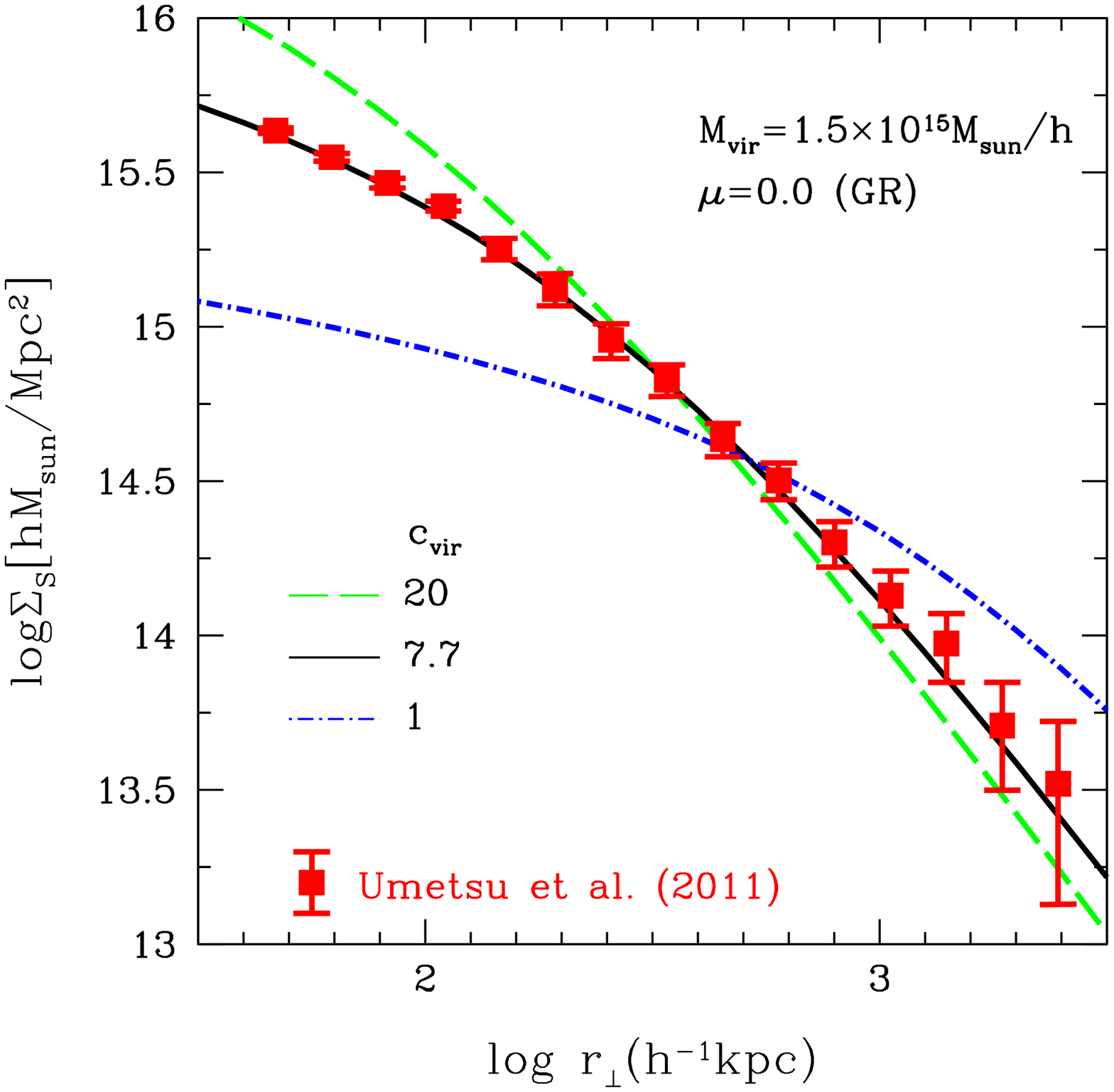}
      \end{center}
    \end{minipage}
    \begin{minipage}{0.5\textwidth}
      \begin{center}
\includegraphics[width=6.2cm,height=6.2cm,clip]{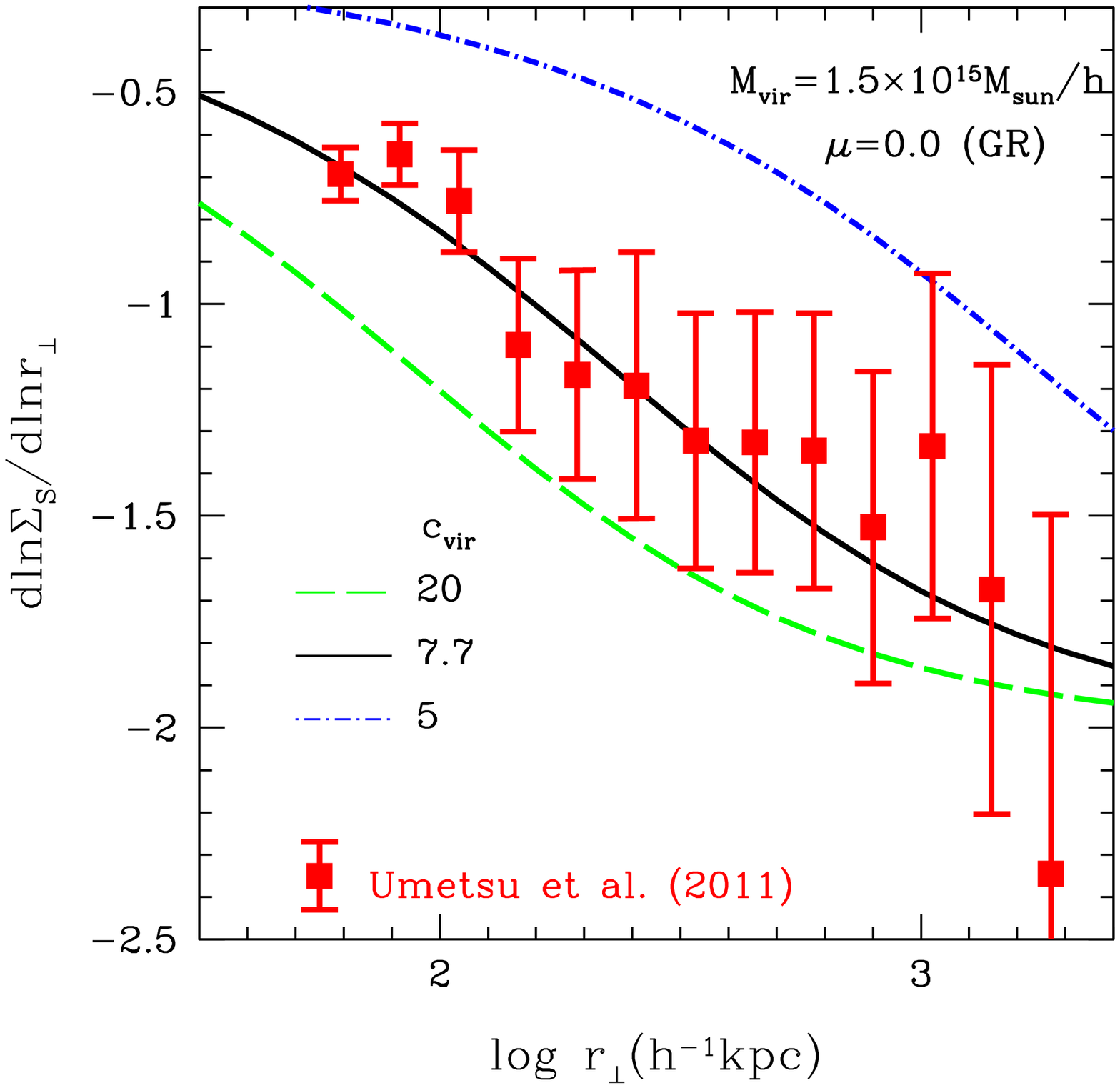}
      \end{center}
    \end{minipage}
  \end{tabular}
\caption{
Same figure as figure \ref{fig:Sigma.Mvir.demo} but with different values 
$c_{\rm vir}=20$ (green dashed curve), $c_{\rm vir}=7.7$ (black solid curve),
and $c_{\rm vir}=1.0$ (blue dot-dashed curve), with 
$M_{\rm vir}=1.5\times10^{15}\solM/h$ fixed. 
Here we assumed Newtonian gravity $\mu=0$ and the NFW profile. 
}
\label{fig:Sigma.cvir.demo}
\end{figure}

%gNFW gammas demo 
%=========================================================================
\begin{figure}[htbp]
  \begin{tabular}{cc}
    \begin{minipage}{0.5\textwidth}
      \begin{center}
\includegraphics[width=6.2cm,height=6.2cm,clip]{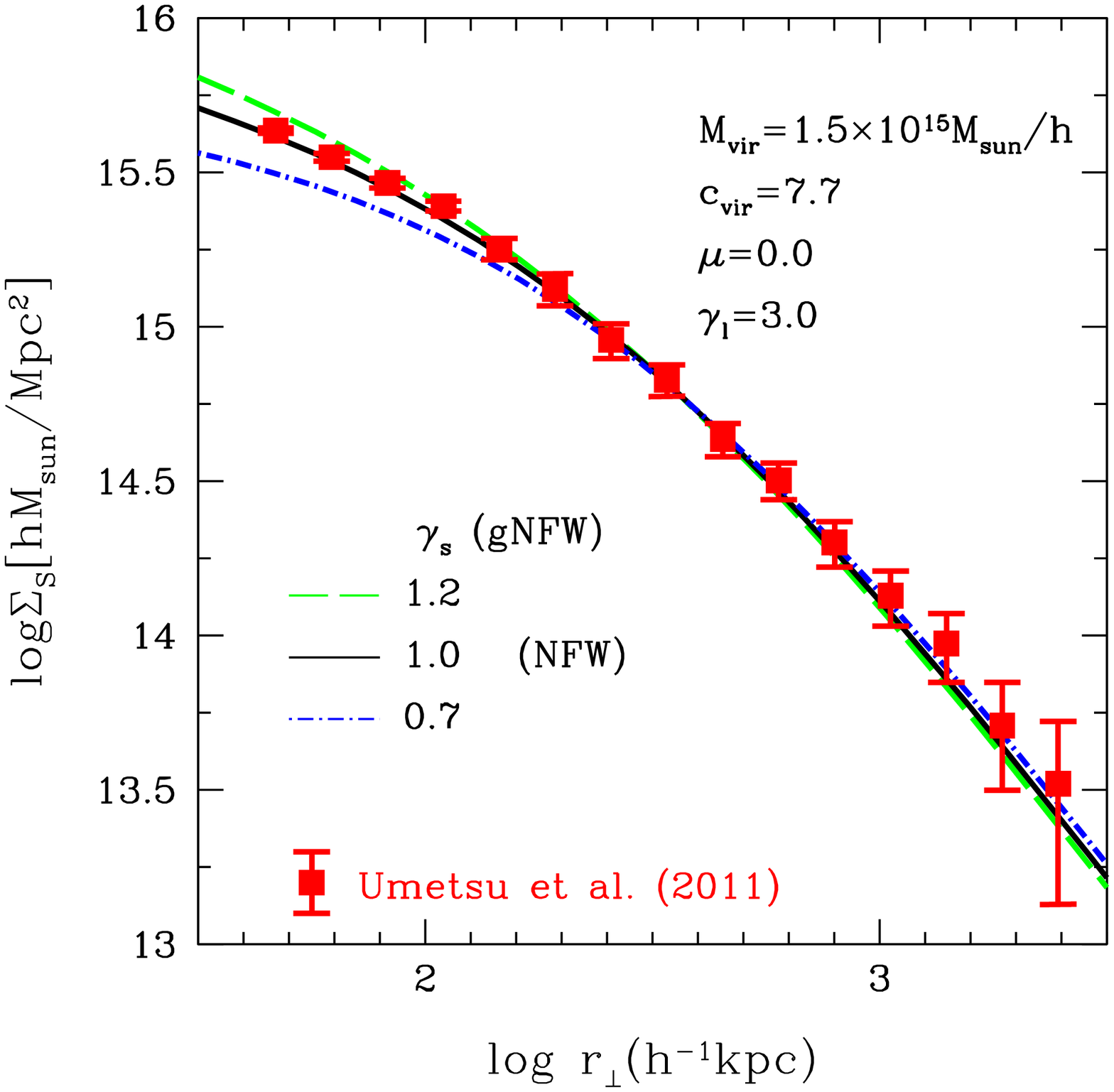}
      \end{center}
    \end{minipage}
    \begin{minipage}{0.5\textwidth}
      \begin{center}
\includegraphics[width=6.2cm,height=6.2cm,clip]{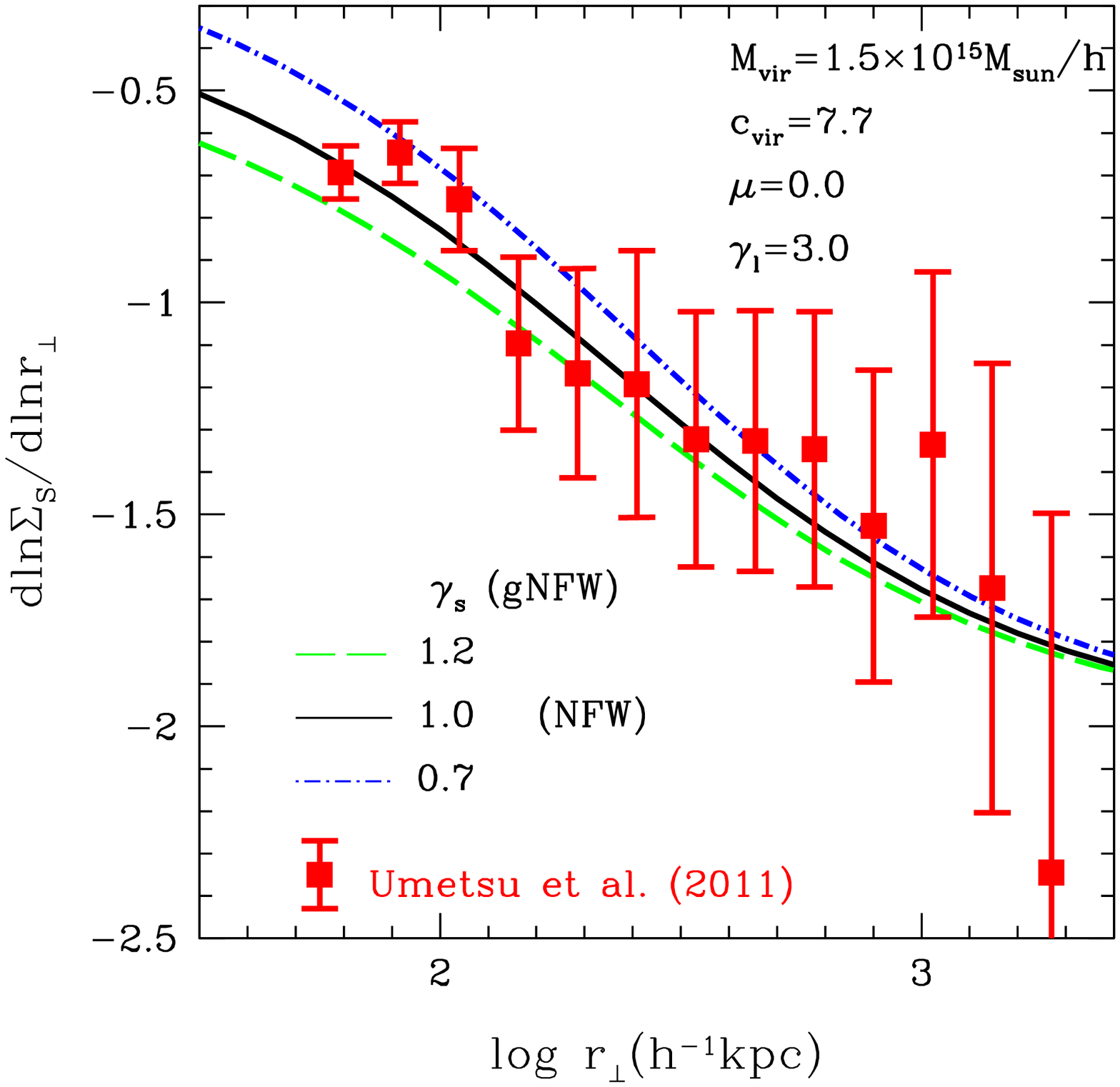}
      \end{center}
    \end{minipage}
  \end{tabular}
\caption{
Same figure as figure \ref{fig:Sigma.Mvir.demo} but for the gNFW profile
with $\gamma_s=1.2$ (green dashed curve), $\gamma_s=1$ (black solid curve), 
and $\gamma_s=0.7$ (blue dot-dashed curve), respectively, with
$\gamma_l=3$ fixed. Note that $\gamma_s=1$ is the NFW profile. 
The other parameters are fixed as 
$M_{\rm vir}=1.5\times10^{15}\solM/h$, $c_{\rm vir}=7.7$, and $\mu=0$.
}
\label{fig:Sigma.gNFW.gammas.demo}
\end{figure}

%gNFW gammal demo 
%=========================================================================
\begin{figure}[htbp]
  \begin{tabular}{cc}
    \begin{minipage}{0.5\textwidth}
      \begin{center}
\includegraphics[width=6.2cm,height=6.2cm,clip]{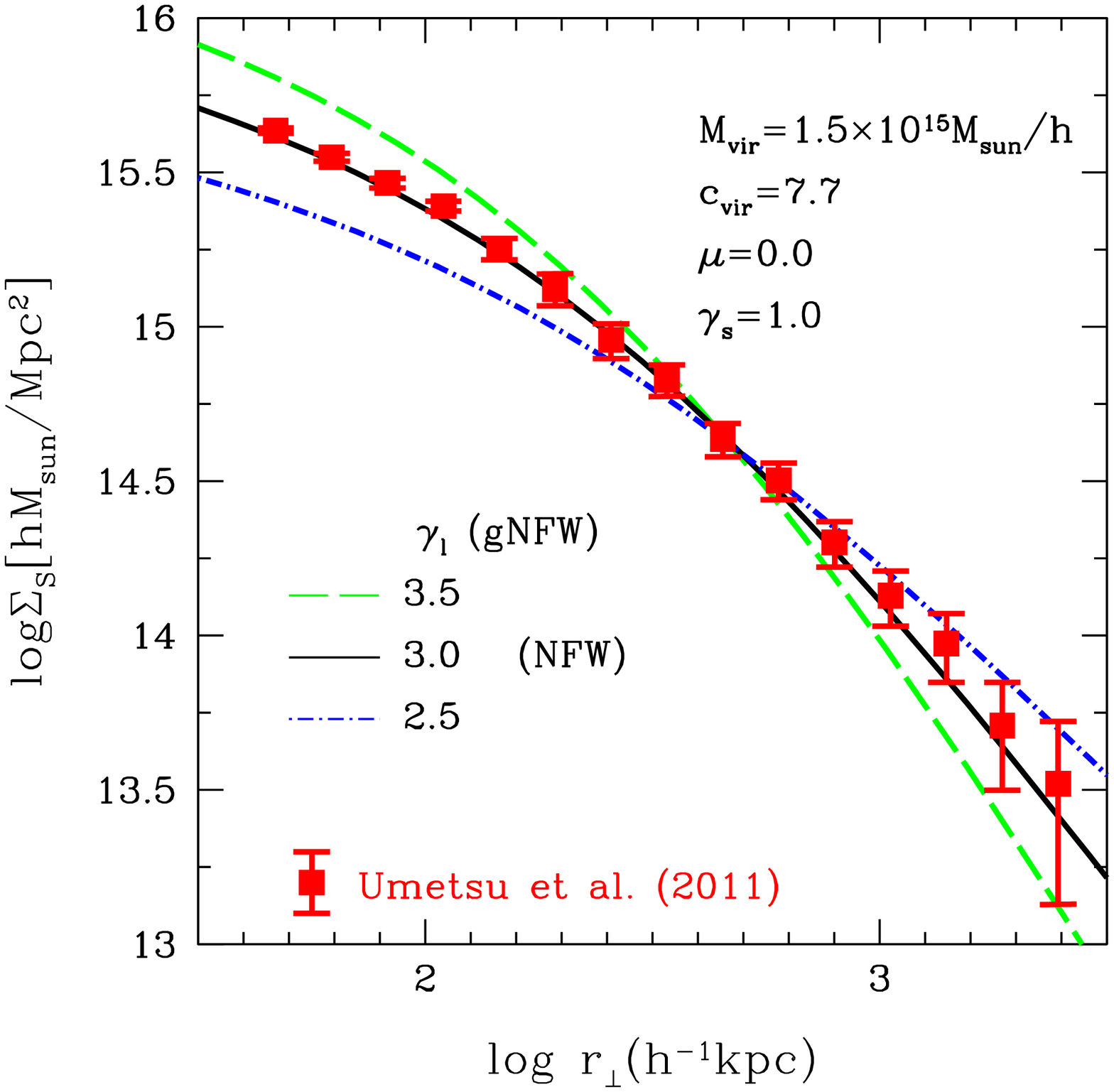}
      \end{center}
    \end{minipage}
    \begin{minipage}{0.5\textwidth}
      \begin{center}
\includegraphics[width=6.2cm,height=6.2cm,clip]{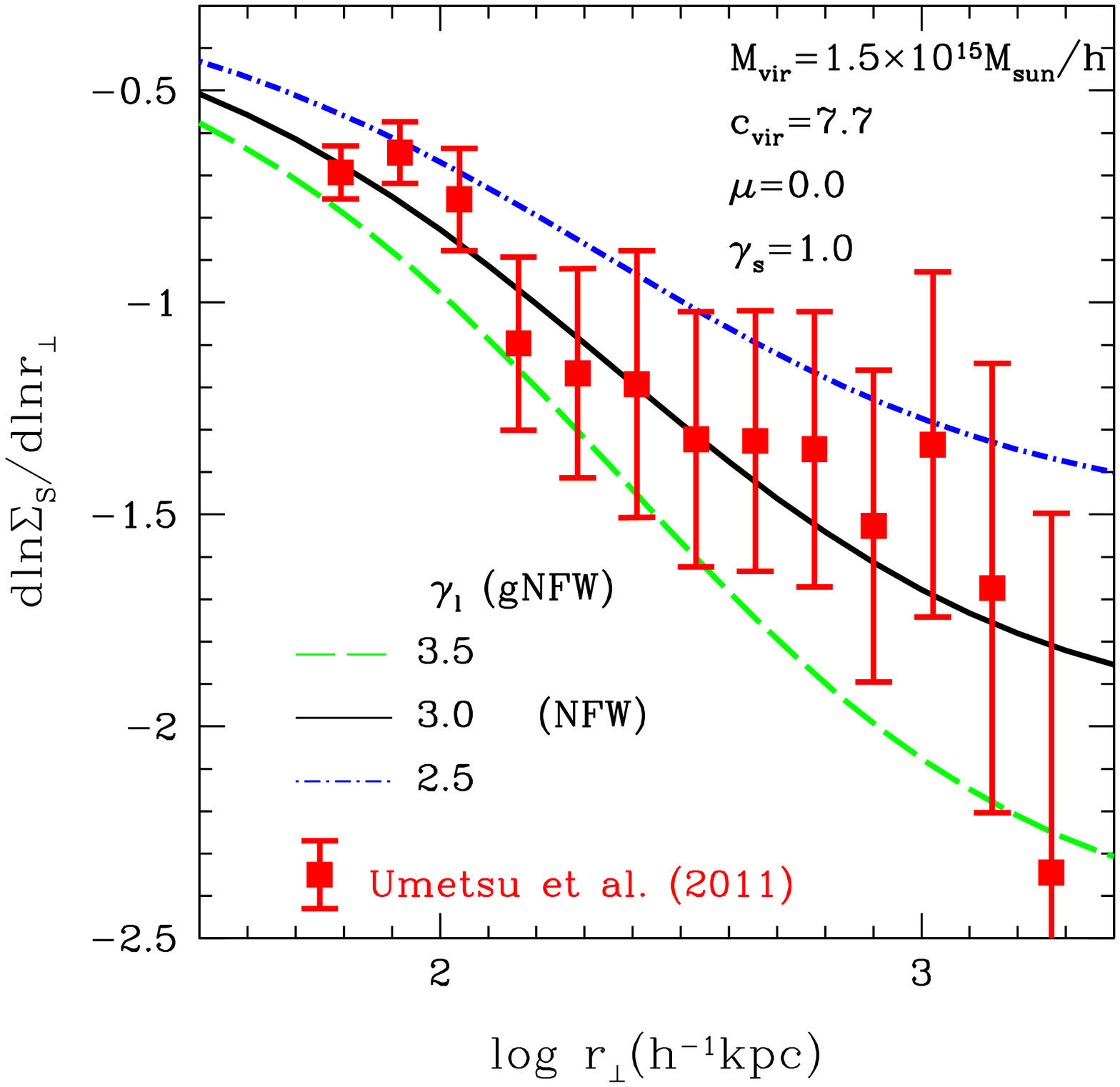}
      \end{center}
    \end{minipage}
  \end{tabular}
\caption{
Same figure as figure \ref{fig:Sigma.Mvir.demo} but for the gNFW profile
with $\gamma_l=3.5$ (green dashed curve), $\gamma_l=3$ (black solid curve), 
and $\gamma_l=2.5$ (blue dot-dashed curve), respectively, with
$\gamma_s=1$ fixed. Note that $\gamma_l=3$ is the NFW profile. 
The other parameters are fixed as 
$M_{\rm vir}=1.5\times10^{15}\solM/h$, $c_{\rm vir}=7.7$, and $\mu=0$.
}
\label{fig:Sigma.gNFW.gammal.demo}
\end{figure}

%Einasto Gamma demo
%=========================================================================
\begin{figure}[htbp]
  \begin{tabular}{cc}
    \begin{minipage}{0.5\textwidth}
      \begin{center}
\includegraphics[width=6.2cm,height=6.2cm,clip]{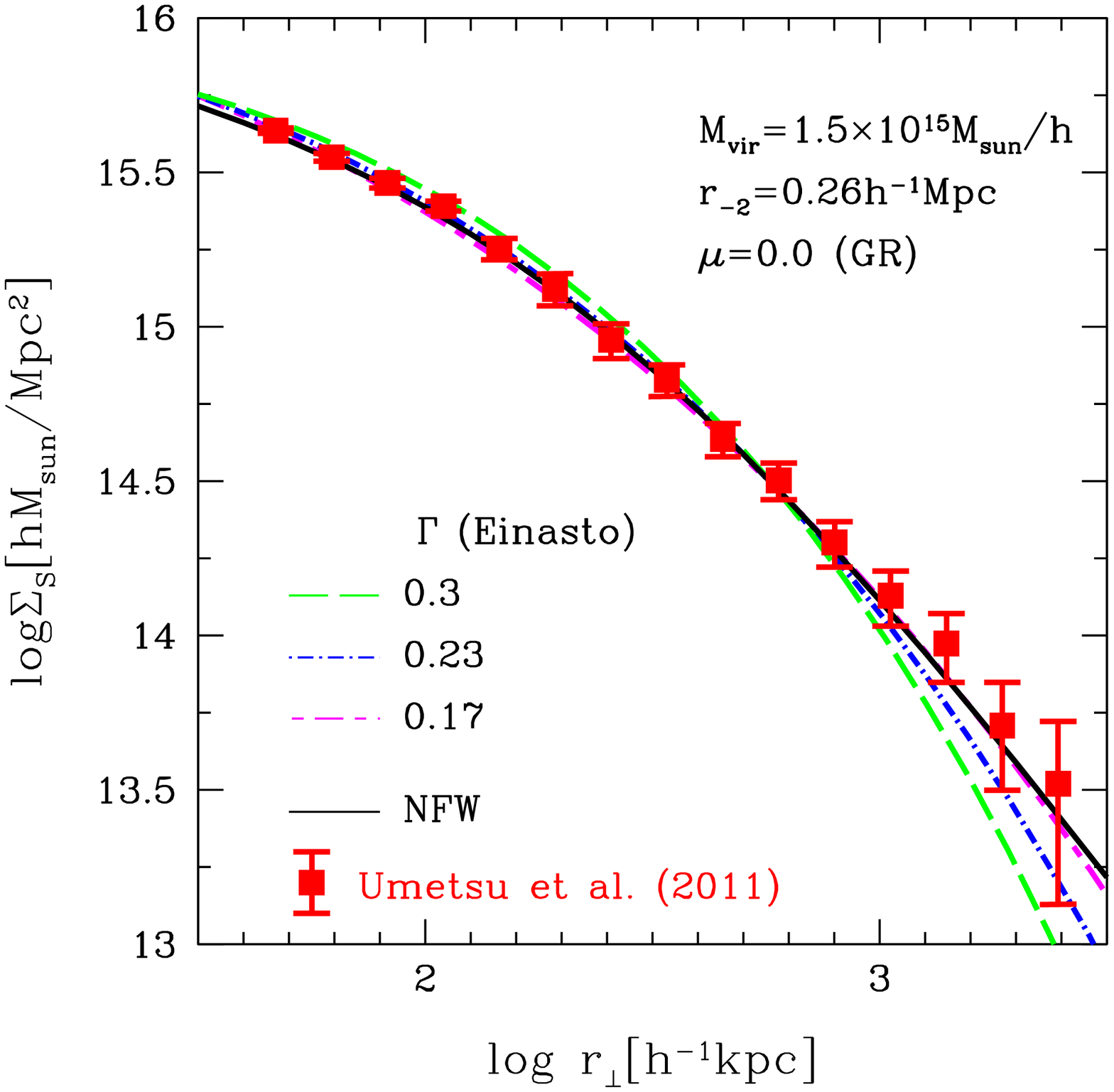}
      \end{center}
    \end{minipage}
    \begin{minipage}{0.5\textwidth}
      \begin{center}
\includegraphics[width=6.2cm,height=6.2cm,clip]{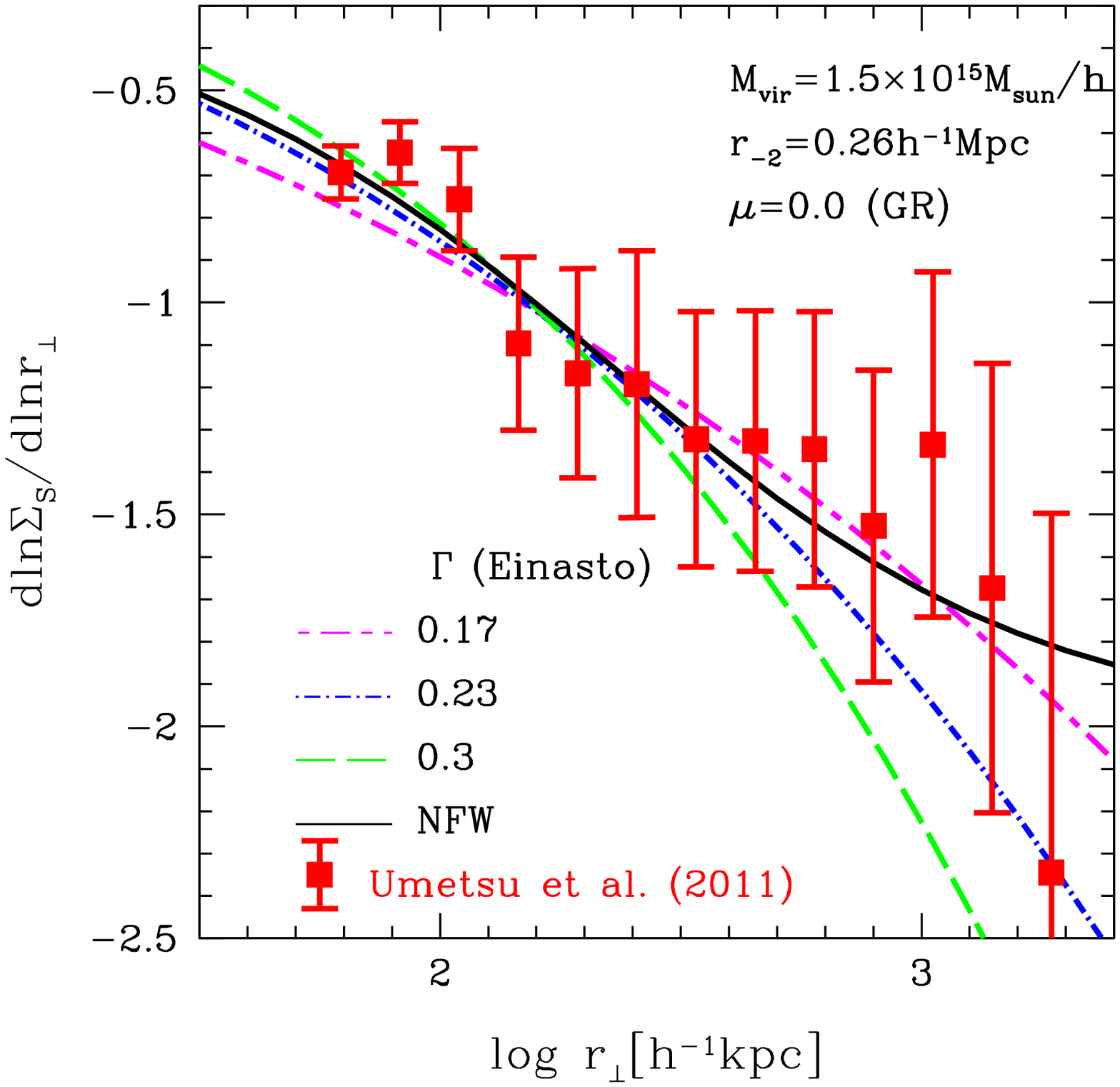}
      \end{center}
    \end{minipage}
  \end{tabular}
\caption{
Same figure as figure \ref{fig:Sigma.Mvir.demo} but for the Einasto profile
with $\Gamma=0.3$ (green dashed curve), $\Gamma=0.23$ (blue dot short-dashed curve), 
and $\Gamma=0.17$ (magenta dot long-dashed curve), and with 
$r_{-2}=0.26 h^{-1}{\rm Mpc}$ fixed. 
The mass parameter is fixed $M_{\rm vir}=1.5\times10^{15}\solM/h$, 
and we assumed Newtonian gravity $\mu=0$. 
For comparison, we also plot the NFW profile (black solid curve). 
}
\label{fig:Sigma.Einasto.Gamma.demo}
\end{figure}

%Einasto r_-2 demo
%=========================================================================
\begin{figure}[htbp]
  \begin{tabular}{cc}
    \begin{minipage}{0.5\textwidth}
      \begin{center}
\includegraphics[width=6.2cm,height=6.2cm,clip]{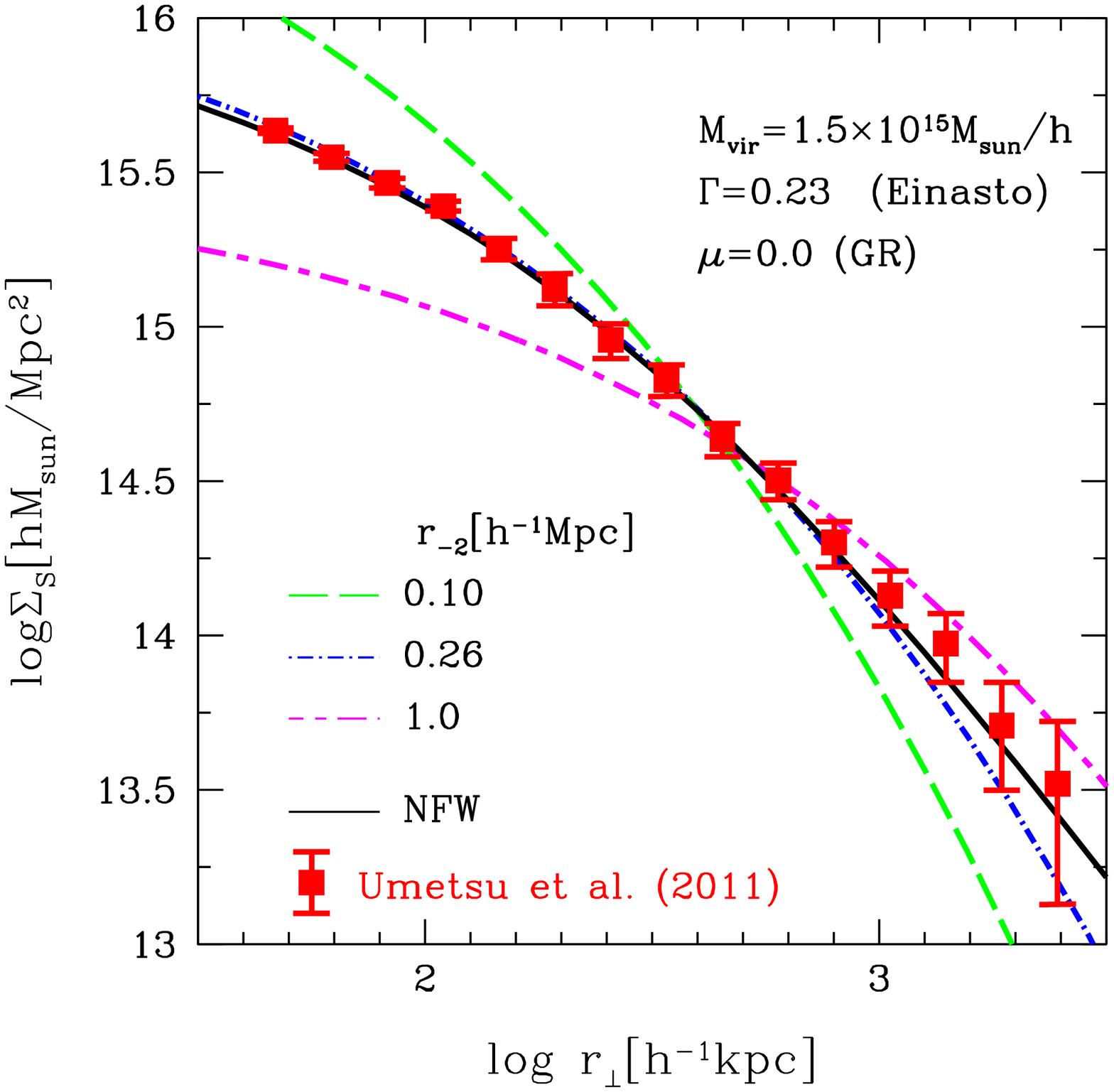}
      \end{center}
    \end{minipage}
    \begin{minipage}{0.5\textwidth}
      \begin{center}
\includegraphics[width=6.2cm,height=6.2cm,clip]{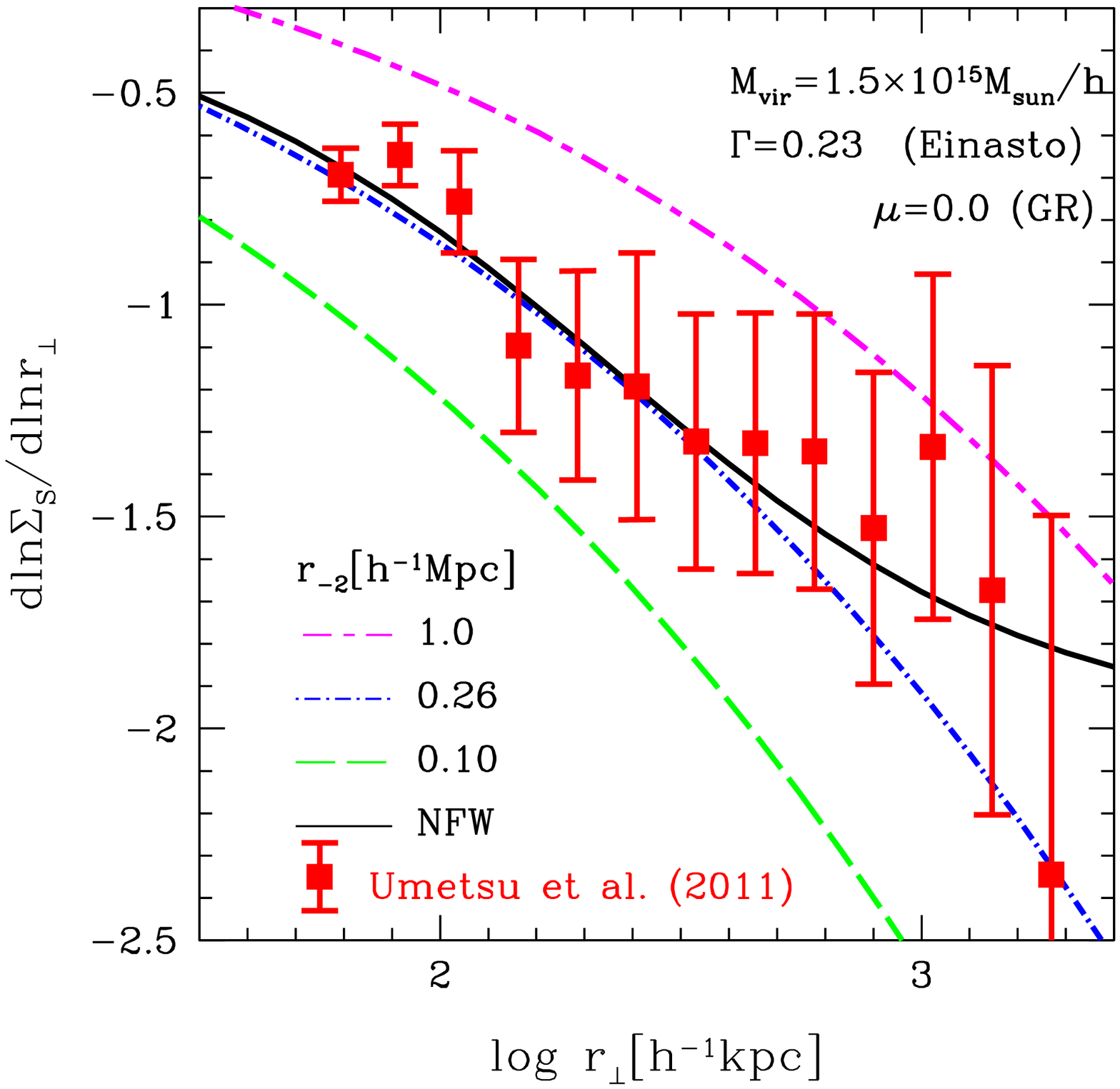}
      \end{center}
    \end{minipage}
  \end{tabular}
\caption{
Same figure as figure \ref{fig:Sigma.Mvir.demo} but for the Einasto profile
wit $r_{-2}=0.1h^{-1}{\rm Mpc}$ (green dashed curve), 
$0.22h^{-1}{\rm Mpc}$ (blue dot short-dashed curve), and 
$1.0h^{-1}{\rm Mpc}$ (magenta dot long-dashed  curve), respectively, 
with $\Gamma=0.23$ fixed. 
The mass parameter is fixed $M_{\rm vir}=1.5\times10^{15}\solM/h$, 
and we assumed Newtonian gravity $\mu=0$. The black solid curve
is the NFW profile. 
}
\label{fig:Sigma.Einasto.r_2.demo}
\end{figure}

%Oguri mu & epsilon & Mvir & cvir demo
%=========================================================================
\begin{figure}
\center \includegraphics[width=32pc]{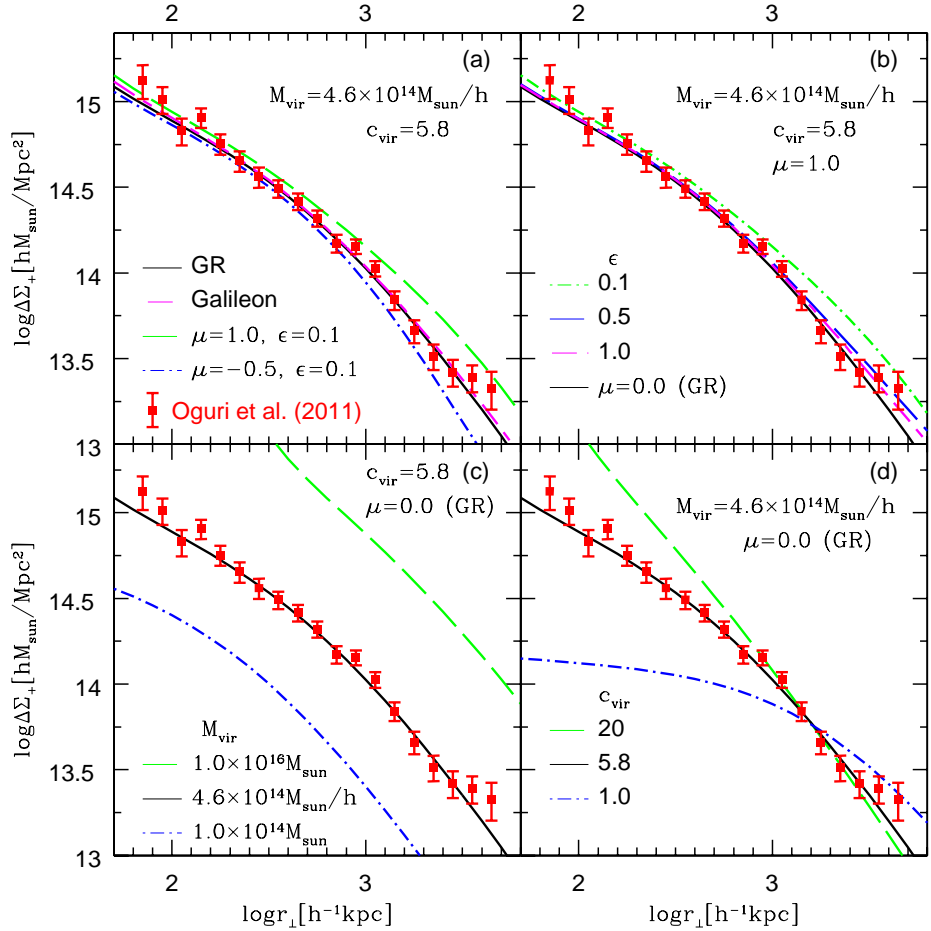}
\caption{
Differential surface mass density $\Delta\Sigma_+(r_\perp)$ in 
as a function of $r_\perp$. The data with the error bar is the result by 
Oguri et al.~\cite{Oguri2011}. 
In the panel (a), the curves adopt the different values of 
$\mu=1.0$ (green dashed curve), 
$\mu=-0.5$ (blue dot short-dashed curve), and $\mu=0$ (black solid curve), 
respectively, with $\epsilon=0.1$ fixed. In this panel the NFW profile 
with $M_{\rm vir}=4.6\times10^{14}\solM/h$ and $c_{\rm vir}=5.8$ is adopted. 
The panel (b) is the same as the panel (a), but we adopted the 
different values of $\epsilon=0.1$ (green dashed curve), $\epsilon=0.5$ 
(blue dot short-dashed curve), and $\epsilon=1.0$ (magenta long-dashed curve), 
respectively, with $\mu=1.0$ fixed. The other parameters of 
$M_{\rm vir}$ and $c_{\rm vir}$ are the same as those of the panel (a). 
The panel (c) is the same as the panel (a), but we adopted the 
different values of $M_{\rm vir}=1.0\times10^{16}\solM$ (green dashed curve),
$4.6\times10^{14}\solM/h$ (black solid curve), and 
$1.0\times10^{14}\solM$ (blue dot-dashed curve), respectively, 
for the NFW profile.
The other parameters are fixed as $\mu=0$ and $c_{\rm vir}=5.8$.
The panel (d) is the same as the panel (c), but adopted the different 
values of $c_{\rm vir}=20$ (green dashed curve), $5.8$ (black solid curve), 
and $1.0$ (blue dot-dashed curve), respectively.
The other parameters are fixed as $M_{\rm vir}=4.6\times 10^{14}\solM$ and 
$\mu=0.0$.
}
\label{fig:DeltaSigma.m.e.M.c.demo}
\end{figure}

%Oguri gNFW & Einasto demo
%=========================================================================
\begin{figure}
\center \includegraphics[width=32pc]{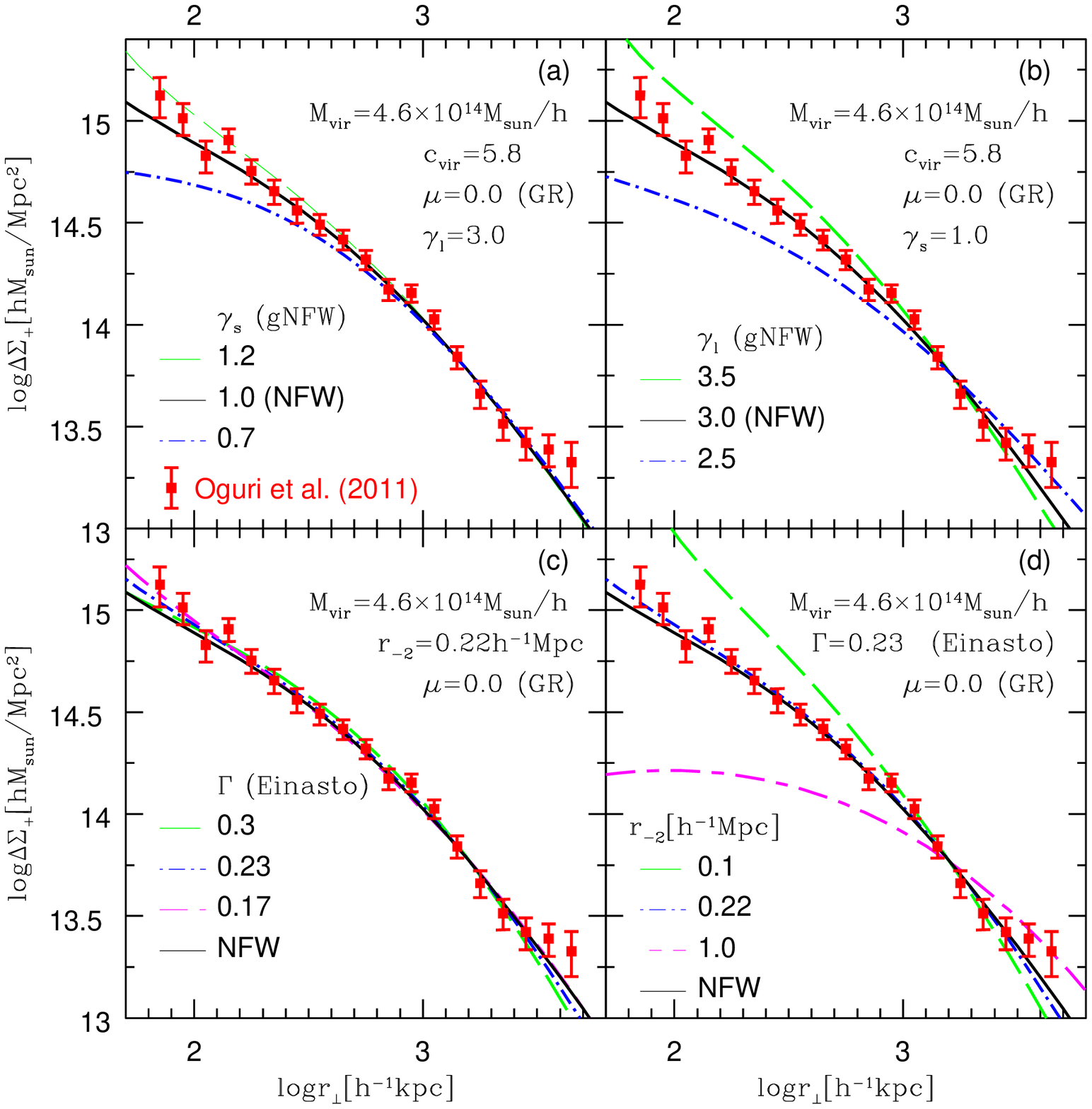}
\caption{Same figure as figure \ref{fig:DeltaSigma.m.e.M.c.demo} but with
different theoretical models.
The panels (a) and (b) assume the gNFW profile, and 
(c) and (d) do the Einasto profile, respectively. 
The panel (a) adopted the gNFW profile with 
$\gamma_s=1.2$ (green dashed curve), $\gamma_s=0.7$ (blue dot-dashed curve), 
and $\gamma_s=1$ (black solid curve), respectively. The other parameters
are fixed as $\gamma_l=3$, 
$M_{\rm vir}=4.6\times10^{14}\solM/h$, $c_{\rm vir}=5.8$, and $\mu=0$. 
In this panel $\gamma_s=1$ is equivalent to the NFW profile. 
The panel (b) is the same as the panel (a), but with 
$\gamma_l=3.5$ (green dashed curve), $\gamma_l=2.5$ (blue dot-dashed curve), 
and $\gamma_l=3$, respectively. In this panel, we fixed $\gamma_s=1$, then
$\gamma_l=3$ is equivalent to the NFW profile.
The panel (c) adopted the Einasto profile, with 
$\Gamma=0.3$ (green dashed curve), $\Gamma=0.23$ (blue dot short-dashed curve), 
and $\Gamma=0.17$ (magenta dot long-dashed curve), respectively. 
The other parameters are fixed $r_{-2}=0.22h^{-1}{\rm Mpc}$,
$M_{\rm vir}=4.6\times10^{14}\solM/h$, and $\mu=0$.
The panel (d) is the same as the panel (c), but with 
$r_{-2}=0.1h^{-1}{\rm Mpc}$ (green dashed curve), 
$0.22h^{-1}{\rm Mpc}$ (blue dot short-dashed curve), and 
$1.0h^{-1}{\rm Mpc}$ (magenta dot long-dashed curve), respectively, 
while $\Gamma=0.23$ fixed. 
}
\label{fig:DeltaSigma.gNFW.Einasto.demo}
\end{figure}

%=========================================================================
\section{Halo density profiles and differential surface mass density}
\label{sec:profile}
%=========================================================================
The halo density profile has been investigated as an important consequence
of the cold dark matter paradigm, which is a key component of the universe. 
Lots of works have been done on the halo density profile
with cosmological N-body simulations
% and strong and weak lensing observations in many literatures 
(e.g.~\cite{Moore,Navarro2004,Merrit2005,Merrit2006,Gao,HayashiWhite,Diemand,Springel2008a,Springel2008b}).
The NFW universal profile~\cite{NFW1996,NFW1997} is reported to work well, 
but the generalized NFW (gNFW) profile has also been 
studied~\cite{Zhao96,JingSuto,Mandelbaum06}. Some studies indicate 
that the Einasto profile~\cite{Einasto} better fits the inner 
cusps~\cite{Graham2006,Navarro2010,Ludlow2011}.

For predicting the surface mass density $\Sigma_{\rm S}(r_\perp)$, 
we need to determine the halo density profile $\rho(r)$.
In the present paper, we adopt the halo density profiles described 
in this section,\footnote{
We only consider the contribution of the cluster halo profile itself 
(1-halo term) to $\Sigma_{\rm S}(r_\perp)$, and we neglect the neighboring 
halos (2-halo term), which is only 
important at large $r_\perp$~\cite{Mandelbaum06}.} 
whose validity is suggested from N-body simulations. 
The halo density profiles could be affected by the modification of gravity,
however, we adopt the same profile irrespective of gravity model, 
whose validity is partially suggested by N-body simulations for the 
DGP model and $f(R)$ model~\cite{dynamicalmass,Zhao11}.
In this section, after briefly reviewing cold dark matter (CDM) 
halo density profiles (see e.g.,~\cite{Navarro2010}),
we demonstrate how the {\it observed} surface mass density depends on 
the modified 
gravity. This indicates the potential ability of testing the gravity theory
with the halo density profile.

%=========================================================================
\subsection{NFW profile}
%=========================================================================
We first consider the NFW profile~\cite{NFW1996,NFW1997} 
\begin{eqnarray}
 \rho(r)={\rho_s\over (r/r_s)(1+r/r_s)^2},
\end{eqnarray}
where $\rho_s=4\rho(r_s)$ is the characteristic density,  $r_s$ is 
the characteristic radius where the slope of the density profile changes.
We define the virial cluster mass and the concentration parameter by
\begin{eqnarray}
M_{\rm vir}&=&{4\pi r_{\rm vir}^3 \rho_{\rm cr}(z_l)\Delta_{\rm vir}\over3},
\\
c_{\rm vir}&=&{r_{\rm vir}\over r_s},
\end{eqnarray}
respectively, which can be used as the parameters of the NFW profile 
instead of $\rho_s$ and $r_s$, where $r_{\rm vir}$ is the virial radius, 
$\Delta_{\rm vir}$ is the virial overdensity,
and $\rho_{\rm cr}(z_l)=\rho_{\rm cr,0} H^2(z_l)/H_0^2$ is the critical 
density. We take the value $\Delta_{\rm vir}=120$.
Note that the virial overdensity and the critical density depend on 
the cosmological background, 
which we do not specify explicitly.
However, when $M_{\rm vir}$ and $c_{\rm vir}$ are marginalized over, 
the dependence on the cosmological background is not necessarily
specified because it is absorbed by the redefinition of
$M_{\rm vir}$ and $c_{\rm vir}$. 

Figure \ref{fig:Sigma.Mvir.demo} shows
the surface mass density $\Sigma_{\rm S}(r_\perp)$ (left panel)
and its logarithmic slope $d\ln\Sigma_{\rm S}/d\ln r_\perp$ (right 
panel), where we adopted the different values of $M_{\rm vir}$ with
the other parameters fixed, which are described therein.
Here the concentration parameter is fixed $c_{\rm vir}=7.7$. 
Note that $\mu=0$ means Newtonian gravity.
$M_{\rm vir}$ changes the amplitude of $\Sigma_{\rm S}(r_\perp)$.
Figure \ref{fig:Sigma.cvir.demo} is the same figure as 
\ref{fig:Sigma.Mvir.demo}, but with adopting the different 
values of $c_{\rm vir}$ and fixing the other parameter.
Here we fixed the virial mass as $M_{\rm vir}=1.5\times10^{15}\solM/h$. 
The slope of the surface mass density significantly depends on $c_{\rm vir}$. 

As we will describe in the next section, we compare the theoretical 
model with observations by introducing the chi square statistics 
(see expression (\ref{eq:chi2umetsu})). Within Newtonian gravity, 
the best-fit value of the chi-squared is $5.8$ for $13$ 
degrees of freedom (DOF), 
which we find for $M_{\rm vir}=1.6\times 10^{15}\solM/h$ and $c_{\rm vir}=7.9$.
(see the top line of table 1).

%=========================================================================
\subsection{gNFW profile}
%=========================================================================
%
The generalized parametrization of the NFW model can be written in
the form (e.g.,~\cite{Zhao96,JingSuto,Mandelbaum06})
\begin{eqnarray}
\rho(r)={\rho_s\over (r/r_s)^{\gamma_s}(1+r/r_s)^{\gamma_l-\gamma_s}}. 
\end{eqnarray}
The logarithmic slope of this density profile is defined
\begin{eqnarray}
\gamma_{3D}(r)\equiv-{d\ln \rho(r)\over d\ln r},
\end{eqnarray}
which reduces to $\gamma_s$ for $r\ll r_s$, and to $\gamma_l$ for $r\gg r_s$. 
The NFW profile is reproduced for $(\gamma_s,~\gamma_l)=(1, 3)$.
Cosmological N-body simulations indicate that an inner logarithmic 
slope of the density profile $\gamma_{3D}(r)\simlt1.2$ and an 
asymptotic outer slope $\gamma_{3D}(r)\simgt2.5$~\cite{Navarro2010}.
We introduce the radius $r_{-2}$ at 
which the outer slope is isothermal, i.e., $\gamma_{3D}(r_{-2})=2$.
For the gNFW profile, 
$r_{-2}=r_s(2-\gamma_s)/(\gamma_l-2)$ and the corresponding 
concentration parameter is 
$c_{-2}\equiv r_{\rm vir}/r_{-2}=c_{\rm vir}(\gamma_l-2)/(2-\gamma_s)$.

Figure \ref{fig:Sigma.gNFW.gammas.demo} shows 
$\Sigma_{\rm S}(r_\perp)$ and $d\ln\Sigma_{\rm S}/d\ln r_\perp$ 
adopting different values of $\gamma_s$ with the other parameters 
fixed as $\gamma_l=3.0$, $M_{\rm vir}=1.5\times10^{15}\solM/h$, $c_{\rm vir}=7.7$, 
and $\mu=0$.
The effect of changing $\gamma_s$ appears only inside halo. 
Figure \ref{fig:Sigma.gNFW.gammal.demo} is the same as figure 
\ref{fig:Sigma.gNFW.gammas.demo}, 
but with adopting the different values of $\gamma_l$ with 
the other parameters fixed as $\gamma_s=1.0$, 
$M_{\rm vir}=1.5\times10^{15}\solM/h$, $c_{\rm vir}=7.7$, 
and $\mu=0$. 
The behavior of $\Sigma_{\rm S}(r_\perp)$ is similar to that of figure 
\ref{fig:Sigma.cvir.demo}, which causes a degeneracy in the 
parameter space. 
We find the best-fit value of the chi-squared $\chi^2/{\rm DOF}=5.5/13$
within Newtonian gravity for $(\gamma_s,\gamma_l)=(0.7, 2.8)$,
using the surface mass density data in~\cite{Umetsu2011b}.
This value of the chi-squared is slightly better than the best-fit 
value $\chi^2/{\rm DOF}=5.8/13$ for the NFW profile with 
$(\gamma_s,\gamma_l)=(1, 3)$. 
The best-fit value for the gNFW profile with 
$(\gamma_s,~\gamma_l)=(1.1, 3.1)$ is $\chi^2/{\rm DOF}=6.2/13$ 
within Newtonian gravity (see table 1).

%=========================================================================
\subsection{Einasto profile}
%=========================================================================
Finally, we consider the Einasto profile~\cite{Einasto}
\begin{eqnarray}
\rho(r)=\rho_{-2}\exp \left(-{2\over\Gamma}
\left[\left({r\over r_{-2}}\right)^\Gamma-1\right]\right),
\end{eqnarray}
where $r_{-2}$ and $\rho_{-2}$ are the radius and the density at which 
$\rho(r)\propto r^{-2}$, which means 
$\gamma_{3D}(r_{-2})=2$ and $\rho_{-2}=\rho(r_{-2})$.
The authors in~\cite{Gao} claimed that CDM halos can be more properly 
described by the Einasto profile than the NFW profile, 
using Millennium simulation.  They also claimed that the best-fit
value of $\Gamma$ increases gradually with the increase of the 
virial mass, from $\Gamma\sim0.16$ for galaxy halos to 
$\Gamma\sim0.3$ for the most massive clusters.

Figure \ref{fig:Sigma.Einasto.Gamma.demo} shows 
$\Sigma_{\rm S}(r_\perp)$ and $d\ln\Sigma_{\rm S}/d\ln r_\perp$ with
adopting different values of $\Gamma$ but with the other parameters 
fixed $r_{-2}=0.26 h^{-1}{\rm Mpc}$, 
$M_{\rm vir}=1.5\times10^{15}\solM/h$, and $\mu=0$. 
$\Gamma$ changes the slope and the amplitude of 
$\Sigma_{\rm S}(r_\perp)$ at large radii.
%when $r_{-2}$ fixed.
%$\Sigma_{\rm S}(r_\perp)$ decreases with an increase in $\Gamma$ for large radii.
%
Figure \ref{fig:Sigma.Einasto.r_2.demo} is the same as 
figure  \ref{fig:Sigma.Einasto.Gamma.demo} but with 
varying $r_{-2}$ with the other parameters fixed $\Gamma=0.23$, 
$M_{\rm vir}=1.5\times10^{15}\solM/h$, and $\mu=0$. 
The slope and the amplitude of $\Sigma_{\rm S}(r_\perp)$ 
strongly depend on $r_{-2}$.

We obtained the best-fit value of the chi-squared $9.9$, $7.2$, and $9.6$
for $13$ DOF for the Einasto profile with $\Gamma=0.17,~0.23,$ and $0.3$, 
respectively
(see table 1). 
Thus the Einasto profile does not fit the observational data better 
than the NFW profile or the gNFW profile within Newtonian gravity. 

%
%
%=========================================================================
% Oguri demo
\subsection{Differential surface mass density}
%=========================================================================
Oguri et al.~\cite{Oguri2011} obtained the differential surface mass 
density, which is defined by
\begin{eqnarray}
\Delta\Sigma_+(r_\perp)\equiv\Sigma_{\rm crit}g_+(r_\perp),
\end{eqnarray}
where $\Sigma_{\rm crit}$ is defined by (\ref{sigmacrit}), 
$g_+(r_\perp)$ is the reduced shear
\begin{eqnarray}
g_+(r_\perp)\equiv{\gamma_+(r_\perp)\over1-\kappa(r_\perp)},
\end{eqnarray}
and the tangential shear is defined by
\begin{eqnarray}
\gamma_+(r_\perp)=\bar{\kappa}(<r_\perp)-\kappa(r_\perp)
={\bar{\Sigma}_{\rm S}(<r_\perp)-\Sigma_{\rm S}(r_\perp)\over\Sigma_{\rm crit}},
\end{eqnarray}
with
\begin{eqnarray}
\bar{\Sigma}_{\rm S}(<r_\perp)={2\over r_\perp^2}\int_0^{r_\perp} dr_\perp'r_\perp'\Sigma_{\rm S}(r_\perp').
\end{eqnarray}

In figures \ref{fig:DeltaSigma.m.e.M.c.demo} and 
\ref{fig:DeltaSigma.gNFW.Einasto.demo}, we demonstrate 
the behavior of the differential surface mass density 
$\Delta\Sigma_+(r_\perp)$ as a function of $r_\perp$, 
comparing with the data in~\cite{Oguri2011}.  
The panels (a), (b), (c) and (d) of figure \ref{fig:DeltaSigma.m.e.M.c.demo}
show how $\Delta\Sigma_+(r_\perp)$ depends on $\mu$, $\epsilon$, 
$M_{\rm vir}$ and $c_{\rm vir}$, respectively, adopting the NFW profile. 
The panels (a) and (b) of figure \ref{fig:DeltaSigma.gNFW.Einasto.demo}
show the dependence on $\gamma_s$ and 
$\gamma_l$ in the gNFW profile, 
while the panels (c) and (d) of figure \ref{fig:DeltaSigma.gNFW.Einasto.demo}
show the dependence on $\Gamma$ and $r_{-2}$ in the Einasto profile, 
respectively.
The behaviors are very similar to those of the surface mass density 
described in the above subsections.

Similar to the surface mass density, we compare the theoretical 
prediction with the observational data by introducing the chi-squared, 
(\ref{eq:chi2oguri}). 
We obtained the best-fit values of chi-squared for the NFW profile, the 
gNFW profile and the Einasto profile within Newtonian gravity as well
as allowing the modified gravity, which are summarized in table 2. 
We find the best-fit value of the chi-squared $\chi^2/{\rm DOF}=12.9/17$ 
for the NFW 
profile within Newtonian gravity, with the best-fit parameter 
$M_{\rm vir}=6.6\times 10^{14}\solM/h$ and $c_{\rm vir}=6.1$. 
This is consistent with the result by Oguri et al.~\cite{Oguri2011}.
The best-fit parameter is given 
$\bar M_{\rm vir}=4.6\times 10^{14}\solM/h$ and $\bar c_{\rm vir}=5.7$,
when we adopt the same definition for the virial mass 
$\bar M_{\rm vir}=4\pi r_{\rm vir}^3\Delta(z)\rho_m(z)$. 

The result is similar to that using the surface mass density in the 
point that the Einasto profile with large value of $\Gamma$
does not better fit the observational data than the NFW profile
and the gNFW profile within Newtonian gravity. 
Thus the Einasto profile with $\Gamma=0.3$ is not favored
within Newtonian gravity. 
%================================================================
%================================================================
% chi2 Sigma gNFW Einasto
%================================================================
\begin{table*}[ttbp]
\begin{center}
% \begin{tabular}{ c c c }
  \begin{tabular}{l c | r l r | r l r} \hline\hline
model & & Newtonian&gravity & & Modified & gravity & \\ \hline
 & $(\gamma_s,~\gamma_l)$ & $M_{\rm vir}[\solM]$ & $c_{\rm vir}(c_{-2})$ & $\chi^2_{\rm GR}$ & $M_{\rm vir}[\solM]$ & $c_{\rm vir}(c_{-2})$ & $\chi^2_{\rm MG}$\\ \hline
{\rm ~NFW} & (1,~3)     & $2.2\times10^{15}$ & 7.9  & 5.8 & $2.2\times10^{15}$ & 8.0  & 5.7\\
{\rm gNFW} & (0.7,~2.8) & $2.2\times10^{15}$ & 13.9(8.6) & 5.5 & $2.2\times10^{15}$ & 13.9(8.6) & 5.5 \\
{\rm gNFW} & (1.1,~3.1) & $2.2\times10^{15}$ & 6.2(7.6)  & 6.2 & $2.2\times10^{15}$ & 6.3(7.7) & 6.0 \\
\hline
 & $\Gamma$ & $M_{\rm vir}[\solM]$ & $r_{-2}$ & $\chi^2_{\rm GR}$ & $M_{\rm vir}[\solM]$ & $r_{-2}$ & $\chi^2_{\rm MG}$ \\ \hline
{\rm Einasto} & 0.17 & $2.5\times10^{15}$ & 0.30 & 9.9 & $2.9\times10^{15}$ & 0.35 & 8.7\\
{\rm Einasto} & 0.23 & $2.2\times10^{15}$ & 0.28 & 7.2 & $2.1\times10^{15}$ & 0.26 & 6.5\\
{\rm Einasto} & 0.3  & $1.9\times10^{15}$ & 0.26 & 9.6 & $1.6\times10^{15}$ & 0.22 & 5.6\\
\hline\hline
% \end{tabular}
 \end{tabular}
    \caption{Best-fit value of the chi-squared of the surface mass density 
$\Sigma_{\rm S}(r_\perp)$ for the NFW profile, the gNFW profile 
and the Einasto profile. Here we fixed the values of 
$\gamma_s$ and $\gamma_l$ for the gNFW profile,
and $\Gamma$ for the Einasto profile, as described in the table.
$\chi_{\rm GR}^2$ is the best-fit value for 
each halo model when Newtonian gravity is assumed, 
and $M_{\rm vir}$ and $c_{\rm vir}$ or $r_{-2}$ (in unit of $h^{-1}{\rm Mpc}$) 
yield the best-fit value.
$\chi_{\rm MG}^2$ is the best-fit value allowing the modified gravity model. 
The number of degrees of freedom (DOF) is $13$ and $11$ for 
$\chi_{\rm GR}^2$ and $\chi_{\rm MG}^2$, respectively.
}
\label{table:chi2.Sigma}
%\end{table*}
%\end{center}
% chi2 DeltaSigma gNFW Einasto
%================================================================
%\begin{center}
%\begin{table*}[htbp]
% \begin{tabular}{ c c c }
\vspace{1.5cm}
  \begin{tabular}{l c | r l r | r l r} \hline\hline
 model & & Newtonian & gravity & & Modified & gravity & \\ \hline
 & $(\gamma_s,~\gamma_l)$ & $M_{\rm vir}[\solM]$ & $c_{\rm vir}(c_{-2})$ & $\chi^2_{\rm GR}$ & $M_{\rm vir}[\solM]$ & $c_{\rm vir}(c_{-2})$ & $\chi^2_{\rm MG}$ \\ \hline
{\rm ~NFW} & (1,~3)     & $6.8\times10^{14}$ & 6.1  & 12.9 & $5.9\times10^{15}$ & 6.6  & 11.7\\
{\rm gNFW} & (0.7,~2.8) & $6.8\times10^{14}$ & 10.9(6.7) & 13.1 & $6.1\times10^{15}$ & 11.7(7.2) & 12.2 \\
{\rm gNFW} & (1.1,~3.1) & $6.8\times10^{14}$ & 4.8(5.9)  & 12.8 & $5.9\times10^{15}$ & 5.3(6.5)  & 11.5 \\
\hline
 & $\Gamma$ & $M_{\rm vir}[\solM]$ & $r_{\rm -2}$ & $\chi^2_{\rm GR}$ & $M_{\rm vir}[\solM]$ & $r_{\rm -2}$ & $\chi^2_{\rm MG}$ \\ \hline
{\rm Einasto}& 0.17 & $7.0\times10^{14}$ & 0.23 & 12.4 & $6.6\times10^{15}$ & 0.21 & 12.2\\
{\rm Einasto}& 0.23 & $6.8\times10^{14}$ & 0.23 & 14.0 & $5.4\times10^{15}$ & 0.18 & 11.4\\
{\rm Einasto}& 0.3  & $6.7\times10^{14}$ & 0.23 & 19.7 & $4.7\times10^{15}$ & 0.17 & 11.8\\
\hline\hline
 \end{tabular}
    \caption{Same table as table 1 but for 
the differential surface mass density $\Delta\Sigma_+(r_\perp)$.
The number of DOF is $17$ and $15$ for 
$\chi_{\rm GR}^2$ and $\chi_{\rm MG}^2$, respectively.
}
\label{table:chi2.DeltaSigma}
\end{center}
\end{table*}

%

%=========================================================================
\section{Results - Comparison with observations -}
\label{sec:results}
%=========================================================================
In this section we demonstrate the constraint on the modified gravity 
parameters $\mu$ and $\epsilon$ using the observational data. 
We first use the surface mass density $\Sigma_{\rm S}(r_\perp)$ 
and its logarithmic slope $d\ln\Sigma_{\rm S}/d\ln r_{\perp}$, 
obtained by Umetsu et al. through the accurate 
strong- and weak-lensing measurements~\cite{Umetsu2011b}.
%We constrain $\mu$ and $\epsilon$ by applying chi square statistics.
We define the chi-squared for the surface mass density by
\begin{eqnarray}
 \chi^2(\mu,\epsilon,M_{\rm vir},c_{\rm vir}) \equiv \sum_{i,j} 
\left[ \Sigma_{\rm S}^{\rm theo}(r_{\perp i})-\Sigma_{\rm S}^{\rm obs}(r_{\perp i})) \right] 
{\rm Cov}_{ij}^{-1}
\left[ \Sigma_{\rm S}^{\rm theo}(r_{\perp j})-\Sigma_{\rm S}^{\rm obs}(r_{\perp j})) \right] ,
\nonumber\\
\label{eq:chi2umetsu}
\end{eqnarray}
where $\Sigma_{\rm S}^{\rm obs}(r_{\perp i})$ and ${\rm Cov}_{ij}^{-1}$ are the 
observational surface mass density and the inverse matrix of the 
$1\sigma$ covariance matrix, respectively, 
at the $i$-th projected radius $r_{\perp i}$, which are obtained 
by stacking individual full surface mass density of four high-mass 
clusters (A1689, A1703, A370, and Cl0024+17) at an average
redshift $\langle z_l \rangle = 0.32$ 
\cite{Umetsu2011a,Umetsu2011b}, and $\Sigma_{\rm S}^{\rm theo}(r_{\perp i})$ 
is the theoretically predicted surface mass density. 

For comparison, we also compute the chi-squared for the logarithmic slope
$d\ln\Sigma_{\rm S}/d\ln r_{\perp}$, which can be 
defined in the similar way to (\ref{eq:chi2umetsu}),
\begin{eqnarray}
 \chi^2(\mu,\epsilon,c_{\rm vir},M_{\rm vir}) \equiv \sum_{i} 
{\left[ d\ln\Sigma_{\rm S}^{\rm theo}(r_{\perp i})/d\ln r_{\perp}
      -d\ln\Sigma_{\rm S}^{\rm obs}(r_{\perp i})/d\ln r_{\perp} \right] ^2
 \over
 [\Delta d\ln\Sigma_{\rm S}(r_{\perp i})/d\ln r_{\perp}]^2},
\label{eq:chi2umetsud}
\end{eqnarray}
where $d\ln\Sigma_{\rm S}^{\rm theo}(r_{\perp i})/d\ln r_{\perp}$ and 
$d\ln\Sigma_{\rm S}^{\rm obs}(r_{\perp i})/d\ln r_{\perp}$ are the
theoretical value and the observational value, respectively,
and $\Delta d\ln\Sigma_{\rm S}(r_{\perp i})/d\ln r_{\perp}$ is 
the $1$-sigma error of the data (see the right panel of 
figure \ref{fig:Sigma.mu.demo}).

\begin{figure}
\center \includegraphics[width=22pc]{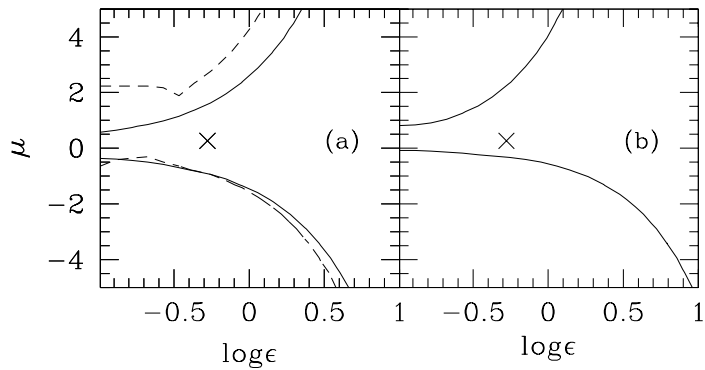}
\caption{
(a) $1$-sigma confidence contour $\Delta\chi^2=2.3$ on the 
$\mu$-$\epsilon$ plane with the surface mass density 
$\Sigma_{\rm S}(r_\perp)$ (solid curve), and with the logarithmic slope 
of the surface mass density $d\ln\Sigma_{\rm S/}/d\ln r_\perp$ 
(dashed curve).
The region sandwiched between a pair of two curves is statistically 
allowed at $1$-sigma level.
The point marked by the cross corresponds 
to the original galileon model (see appendix \ref{sec:Galileon}).
(b)  $1$-sigma confidence contour $\Delta\chi^2=2.3$ on 
$\mu$-$\epsilon$ plane with the differential surface mass density 
$\Delta\Sigma_+(r_\perp)$. In these panels, the NFW profile is assumed.
}
\label{fig:cont.e_m.Umetsu.NFW}
\end{figure}

The left panel of figure \ref{fig:cont.e_m.Umetsu.NFW}
shows the contour of $\Delta \chi^2$,  to show
the relative confidence level of $\mu$ and $\epsilon$ with 
respect to their best-fit values  
\begin{eqnarray}
 \Delta\chi^2(\mu, \epsilon) \equiv 
\chi^2(\mu, \epsilon, M_{\rm vir,local~min}, c_{\rm vir,local~min}) - 
\chi^2(\mu_{\rm min}, \epsilon_{\rm min}, M_{\rm vir,min}, c_{\rm vir,min}).
\label{eq:dchi2}
\end{eqnarray}
%which follows the $\chi^2$ distribution for 2 degree of freedom.
In (\ref{eq:dchi2}), $\mu_{\rm min}$, $\epsilon_{\rm min}$, $M_{\rm vir,min}$, 
and $c_{\rm vir,min}$ denote their best-fit values which globally minimize 
the value of $\chi^2$, while $M_{\rm vir,local~min}$ and $c_{\rm vir,local~min}$ 
are the values that locally minimize the $\chi^2$ for a given set of 
values of $\mu$ and $\epsilon$, where we consider the range
\begin{eqnarray} 
10^{13}\solM\leq &M_{\rm vir}& \leq 10^{16}\solM,\\
0.01\leq &c_{\rm vir}& \leq 40,
\end{eqnarray}
and
\begin{eqnarray} 
0.1\leq &\epsilon& \leq 10,\\
-5.0\leq &\mu& \leq 5.0.
\end{eqnarray}
Our results does not depend on this choice significantly. 
In table \ref{table:chi2.Sigma} we listed the best-fit value of 
the chi-squared of (\ref{eq:chi2umetsu})
$\chi^2_{\rm MG}=\chi^2(\mu_{\rm min}, \epsilon_{\rm min}, $
$M_{\rm vir,min}, c_{\rm vir,min})$, 
for each halo model. 

In the left panel of figure \ref{fig:cont.e_m.Umetsu.NFW}, 
the region sandwiched between the  solid curves is the $1$-sigma 
allowed region with the chi-squared of the surface mass density 
(\ref{eq:chi2umetsu}), while the region between the dashed curves
is the same but with the chi-squared of the logarithmic slope 
(\ref{eq:chi2umetsud}). Here we assumed the NFW profile. 
The small cross in this figure marks the value
of the original galileon model, 
$\mu=0.26$ and $\epsilon=0.53$ 
at the redshift $0.32$ in the left panel.

\begin{figure}
\center \includegraphics[width=22pc]{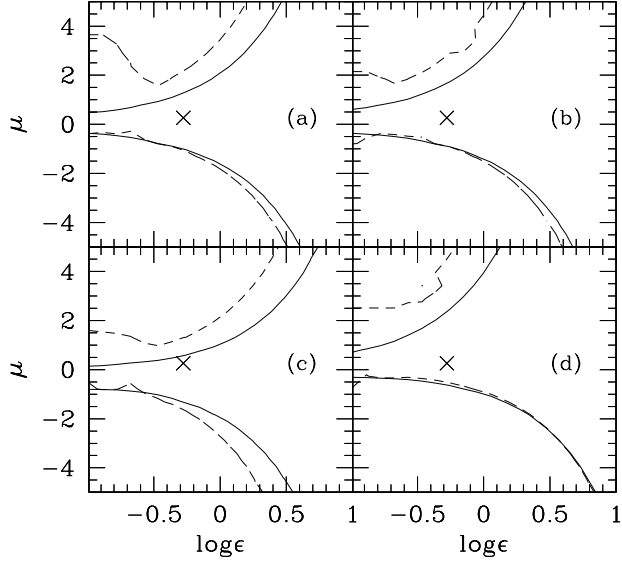}
\caption{Each panel is the same as the left panel of figure 
\ref{fig:cont.e_m.Umetsu.NFW}, but adopted the different halo profile. 
The panels (a) and (b) adopted the gNFW profile with 
$(\gamma_s,\gamma_l)=(0.7, 2.8)$ and $(\gamma_s,\gamma_l)=(1.1, 3.1)$, 
while (c) and (d) adopted the Einasto profile with 
$\Gamma=0.17$ and $\Gamma=0.23$, respectively.
}
\label{fig:cont.Sigma.gNFW.Einasto}
\end{figure}

Note that we adopted the NFW profile in figure \ref{fig:cont.e_m.Umetsu.NFW},
while the panels in figure \ref{fig:cont.Sigma.gNFW.Einasto} assumed the 
different profiles. The panels (a) and (b) adopted the gNFW profile with
$(\gamma_s, \gamma_l)=(0.7, 2.8)$ and $(1.1, 3.1)$, respectively, 
while (c) and (d) adopted the Einasto profile with $\Gamma=0.17$ and $0.23$, 
respectively.

We also investigate the constraint from the data of the differential 
surface mass density $\Delta\Sigma_+(r_\perp)$, obtained by 
Oguri et al.~\cite{Oguri2011}.
We define the chi-squared for $\Delta\Sigma_+(r_\perp)$ by
\begin{eqnarray}
 \chi^2 \equiv \sum_i 
{\left[ \Delta\Sigma_+^{\rm theo}(r_{\perp i})-\Delta\Sigma_+^{\rm obs}(r_{\perp i})) \right]^2 
\over \sigma_{\rm obs}^2(r_{\perp i})},
\label{eq:chi2oguri}
\end{eqnarray}
where $\Delta\Sigma_+^{\rm obs}(r_{\perp i})$ and $\sigma_{\rm obs}(r_{\perp i})$ are 
the {\it observed} differential surface mass density and the $1\sigma$ error
for the $i$-th projected radius $r_{\perp i}$ in~\cite{Oguri2011}, and 
$\Delta\Sigma_+^{\rm theo}(r_{\perp i})$ is the theoretical differential 
surface mass density.
The right panel of figure \ref{fig:cont.e_m.Umetsu.NFW} is the contour 
of $\Delta\chi^2$ for the differential surface mass density, defined 
in the same way as (\ref{eq:dchi2}). 
The region sandwiched between the solid curves is the $1$-sigma 
allowed region, and the point marked by the cross is the original 
galileon model, $\mu=0.19$ and $\epsilon=0.43$ at the 
redshift $z=0.47$. 
In the right panel of figure \ref{fig:cont.e_m.Umetsu.NFW}, we 
adopted the NFW profile, and the best-fit value of the chi 
squared is noted in table \ref{table:chi2.DeltaSigma} as $\chi^2_{\rm MG}$. 

\begin{figure}
\center \includegraphics[width=22pc]{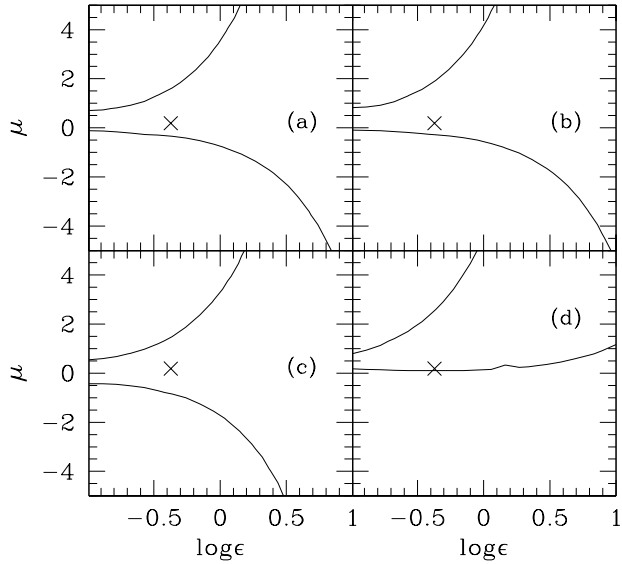}
\caption{Same figure as figure \ref{fig:cont.Sigma.gNFW.Einasto} but
with the differential surface mass density $\Delta\Sigma_{+}(r_\perp)$. 
}
\label{fig:cont.DeltaSigma.gNFW.Einasto}
\end{figure}

The panels in figure \ref{fig:cont.DeltaSigma.gNFW.Einasto} are the
same as the right panel of figure \ref{fig:cont.e_m.Umetsu.NFW}, 
but with adopting the different profiles. 
The panels (a) and (b) of figure \ref{fig:cont.DeltaSigma.gNFW.Einasto} 
assume the gNFW profile with $(\gamma_s$, $\gamma_l)=(0.7, 2.8)$ and 
$(1.1, 3.1)$, while (c) and (d) adopt the Einasto profile with 
$\Gamma=0.17$ and $0.23$, respectively.
Note that the panel (d) indicates that the modified gravity is 
favored than Newtonian gravity when the Einasto profile with 
large $\Gamma$ is assumed. 

% constraints
% Umetsu NFW,  Oguri NFW 
%=========================================================================

% Umetsu gNFW & Einasto constraint
%=========================================================================

% Oguri gNFW & Einasto constraint
%=========================================================================

%=========================================================================
%=========================================================================
\section{Discussion}
\label{sec:discussion}
%=========================================================================

Our main result is symbolically expressed by figure 
\ref{fig:cont.e_m.Umetsu.NFW}.
From figure \ref{fig:cont.e_m.Umetsu.NFW}, 
Newtonian gravity, i.e., $\mu=0$ or $\epsilon\gg1$, is favored, 
and the original galileon model is in the $1$-sigma allowed 
region, which is not rejected. 
One can see that the data put a constraint $|\mu|\simlt 1$ when 
$\epsilon\simlt 1$, but the constraint on $\mu$ becomes weaker 
as $\epsilon$ becomes larger. For instance, $-0.4\simlt\mu\simlt0.6$ 
for $\epsilon=0.1$ and $-1.5\simlt\mu\simlt2.5$ for $\epsilon=1.0$.

The Vainshtein radius $r_V$ depends on $\epsilon$ and 
$M_{\rm vir}$, from eq.~(\ref{eq:Vainshtein}). 
The method in the present paper cannot constrain 
the model in which the Vainshtein radius $r_V$ is larger than the 
cluster scale because the effect of the modified gravity 
becomes very weak inside the Vainshtein radius.
The data we used is limited by $r \simlt3~h^{-1}{\rm Mpc}$ for 
the surface mass density, and by $r \simlt5~h^{-1}{\rm Mpc}$ for 
the differential surface mass density. 
These facts set a limit of constraining $\mu$ and $\epsilon$.
On the other hand, when the Vainshtein radius $r_V$ is smaller 
than the cluster scales $r_{\rm vir}$, 
the effect of the modified gravity appears inside the cluster scale. 
Therefore, we can put a rather tight 
constraint on $\mu$ for small $\epsilon$.

In tables \ref{table:chi2.Sigma} and \ref{table:chi2.DeltaSigma},
respectively, the best-fit value of the chi-squared 
is listed for each halo model, using the surface mass density and 
the differential surface mass density.
$\chi^2_{\rm MG}$ is the best-fit value allowing 
the modified gravity model.
Note that $\chi^2_{\rm MG}$ is improved, 
compared with $\chi^2_{\rm GR}$, which is the best-fit value 
within Newtonian gravity.
We see slight improvement for the NFW profile and the gNFW profile. 
This reflects that Newtonian gravity is favored from the data as long
as the halo profile is the NFW profile or the gNFW profile. 

\begin{figure}[htbp]
  \begin{tabular}{cc}
    \begin{minipage}{0.5\textwidth}
      \begin{center}
\includegraphics[width=6.2cm,height=6.2cm,clip]{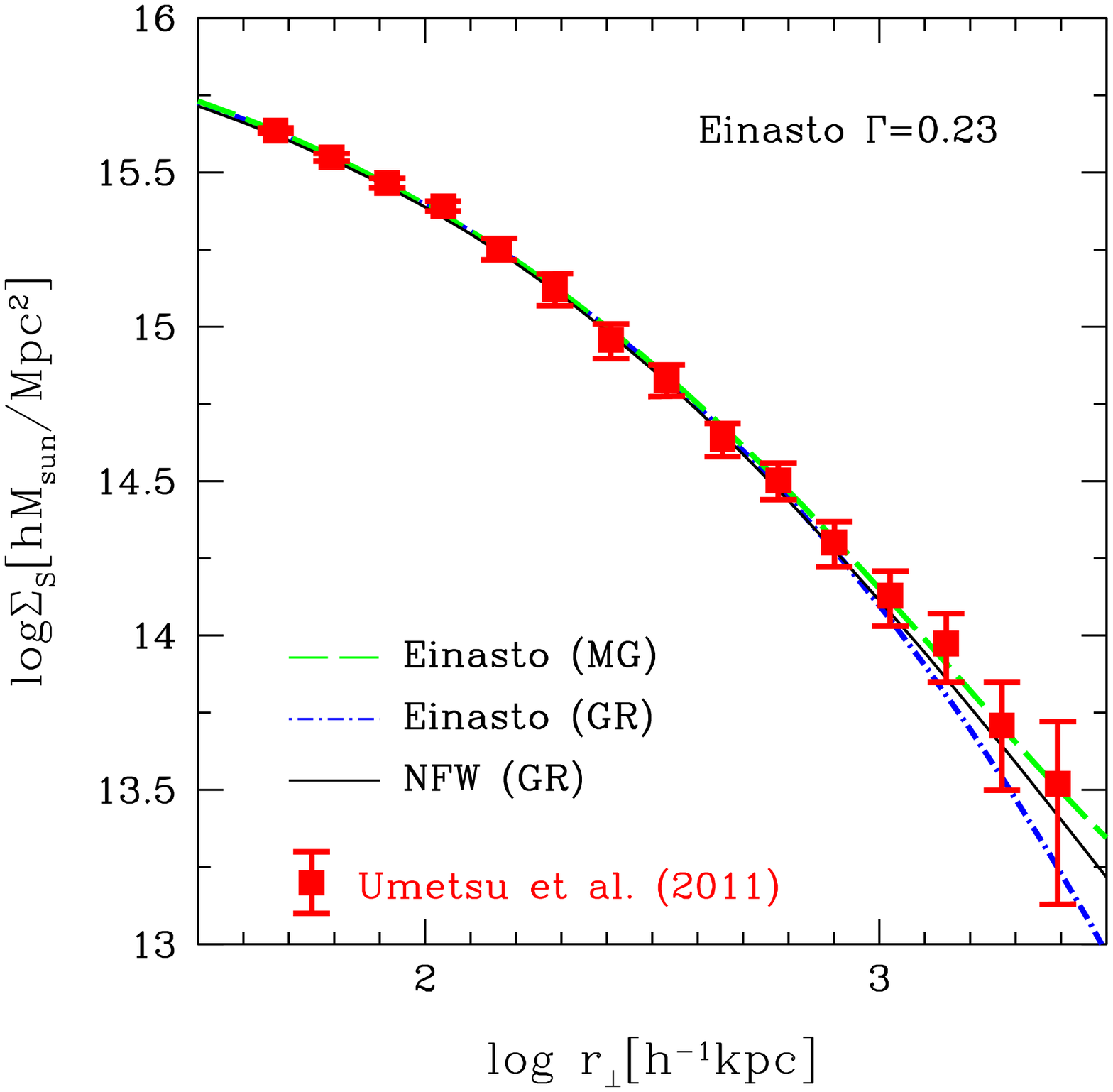}
      \end{center}
    \end{minipage}
    \begin{minipage}{0.5\textwidth}
      \begin{center}
\includegraphics[width=6.2cm,height=6.2cm,clip]{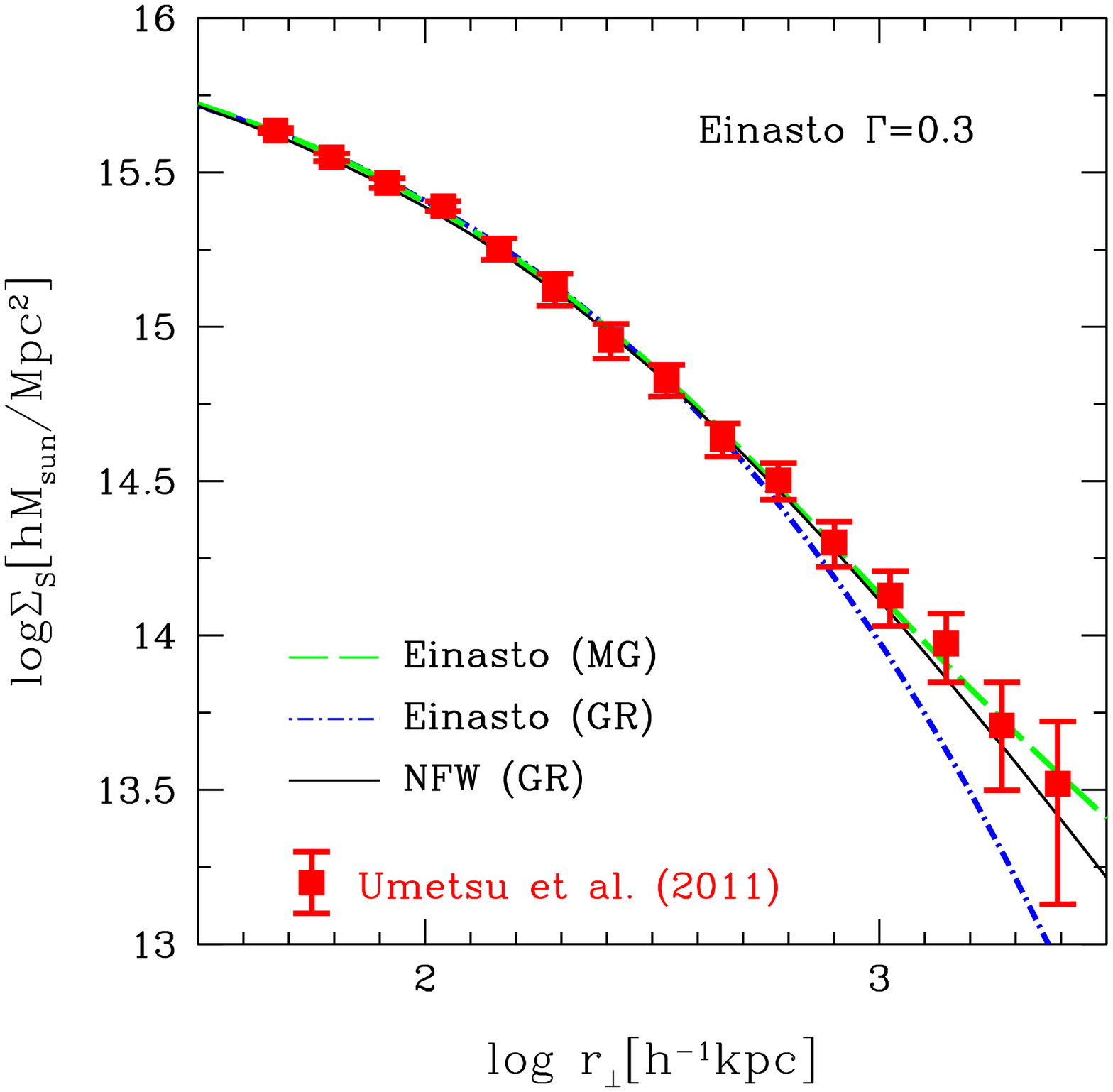}
      \end{center}
    \end{minipage}
  \end{tabular}
\caption{This figure compares the {\it observed} surface mass density 
$\Sigma_{\rm S}(r_\perp)$ and the theoretical curve assuming 
the Einasto profile with $\Gamma=0.23$ (left panel) and 
$\Gamma=0.3$ (right panel), respectively.
In each panel, the blue dot and short-dashed curve is the best-fit 
curve within the Newtonian gravity, while the 
green dashed curve is the best-fit curve of the modified gravity. 
The solid curve assumes the NFW profile within the Newtonian gravity. 
The left (right) panel adopted $\Gamma=0.23 (0.3)$, and the best-fit 
parameters for the modified gravity model are $\mu=5.0$ and $\epsilon=4.5$ 
($\mu=3.4$ and $\epsilon=0.69$), respectively.
}
\label{fig:Sigma.Einasto.fit}
\end{figure}

\begin{figure}[htbp]
  \begin{tabular}{cc}
    \begin{minipage}{0.5\textwidth}
      \begin{center}
\includegraphics[width=6.2cm,height=6.2cm,clip]{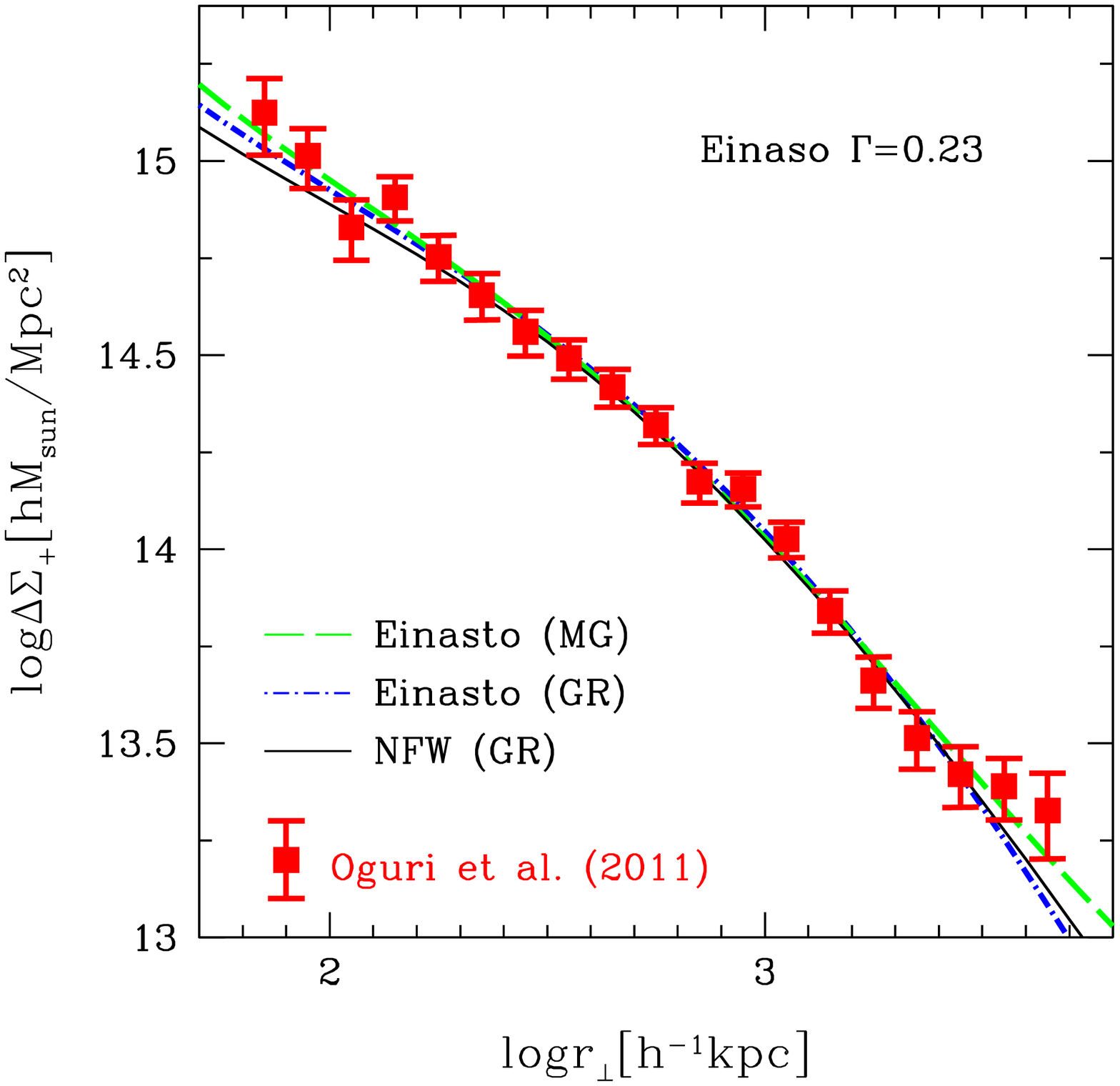}
      \end{center}
    \end{minipage}
    \begin{minipage}{0.5\textwidth}
      \begin{center}
\includegraphics[width=6.2cm,height=6.2cm,clip]{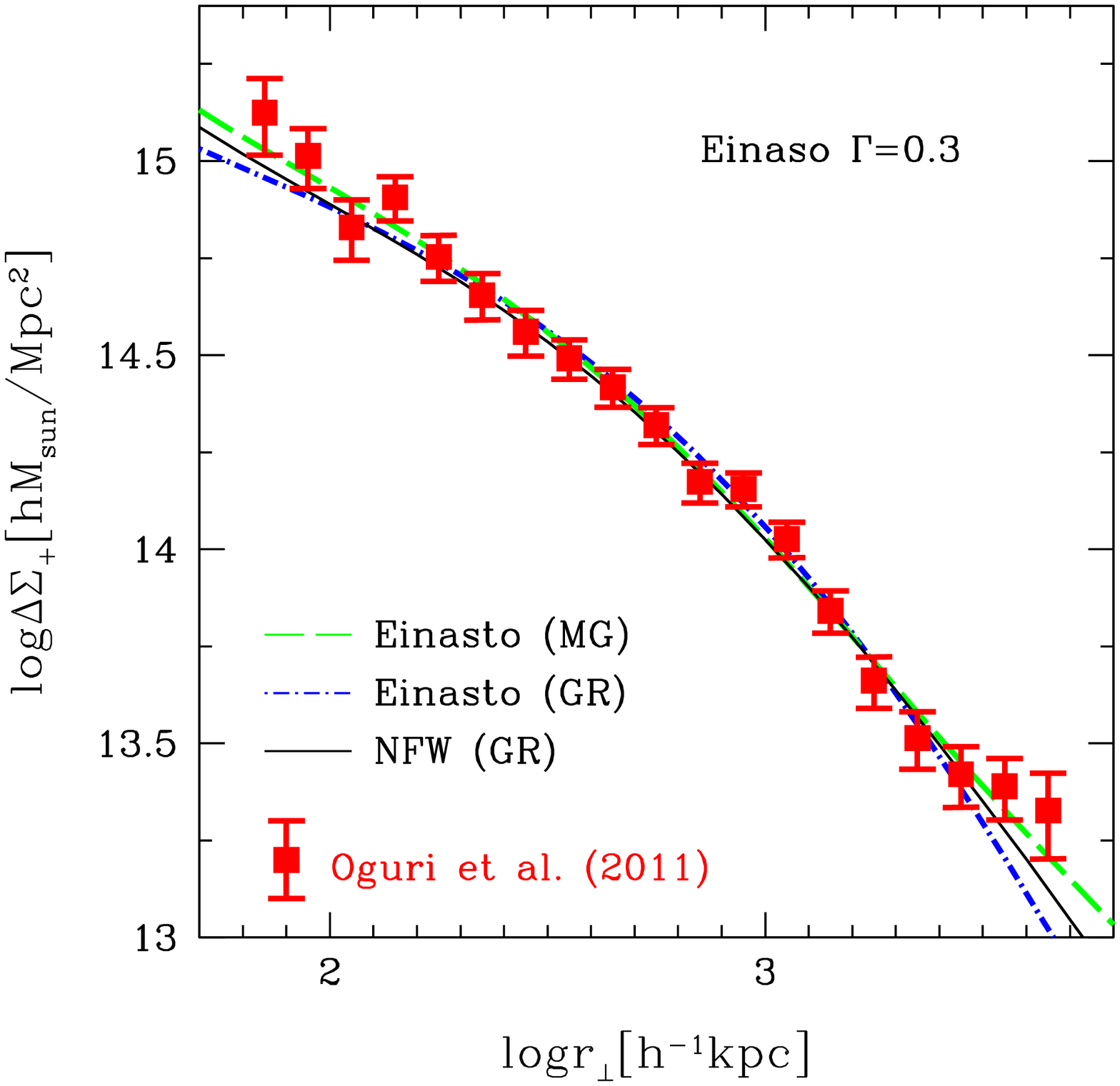}
      \end{center}
    \end{minipage}
  \end{tabular}
\caption{
Same figure as figure \ref{fig:Sigma.Einasto.fit}, but for the 
differential surface mass density $\Delta\Sigma_+(r_\perp)$. 
The best-fit parameter for the modified gravity is 
$\mu=5.0$ and $\epsilon=1.8$ ($\mu=5.0$ and $\epsilon=1.1$) 
in the left (right) panel, respectively.
}
\label{fig:DeltaSigma.Einasto.fit}
\end{figure}

Another point is that the best-fit value is significantly 
improved by allowing the modified gravity
for the Einasto profile with $\Gamma=0.23$ and $\Gamma=0.3$.
This can be explained as follows by using figures 
\ref{fig:Sigma.Einasto.fit} and \ref{fig:DeltaSigma.Einasto.fit}, 
which show how the fitting is improved by the modified gravity model. 
Figure \ref{fig:Sigma.Einasto.fit} shows that the fitting of 
$\Sigma_S(r_\perp)$ is improved at large radii by the modified gravity model. 
Figure \ref{fig:DeltaSigma.Einasto.fit} 
shows that the fitting of $\Delta\Sigma_+(r_\perp)$ is 
improved at large radii as well as at smaller radii. 
Thus the modified gravity model is favored if the halo follows the 
Einasto profile with large value of $\Gamma$. 
This is the reason why the modified gravity is favored in
the panel (d) of figure \ref{fig:cont.DeltaSigma.gNFW.Einasto}.

The above results rely mostly on the behavior of the theoretical curves in
the outer region. Let us check the validity of our theoretical
modeling in the outer region. The behavior in the outer region is
determined by the halo density profile which we choose and the modified
gravity effect of the scalar field. When the density profile is fixed,
the modified gravity's effect of the scalar field is properly taken
into account by using the expression (\ref{defsigmaS}).
In the present paper, however, we assumed the halo density profiles
irrespective of the modified gravity model. This point will need
to be tested carefully, though there exists a supporting evidence
in reference ~\cite{dynamicalmass,Zhao11}.
In the modeling of the halo density profile, we only considered the
contribution of the cluster halo profile itself (1-halo term), and
we neglect the neighboring halos (2-halo term), which will be only
important at large radius ~\cite{Mandelbaum06}.

%=========================================================================

%=========================================================================

%=========================================================================
%=========================================================================
\section{Summary and Conclusion}
\label{sec:summary}
%=========================================================================

We investigated a constraint on the subclass of the generalized 
galileon model endowed with the Vainshtein mechanism, which 
includes a wide class of scalar-tensor gravity theories. 
%for testing a observational consequence 
%in a subclass of general second-order scalar-tensor theories.
The generalized galileon model we considered is characterized by 
the two parameters, $\mu$ and $\epsilon$, which describe the amplitude 
and the typical scale of the modification of gravity of the Vainshtein
mechanism, respectively.
We have investigated how the transition in the Vainshtein mechanism 
in the generalized galileon model appears due to a scalar degree 
of freedom, which leads to a modified relation between the lensing 
potential and the density perturbations of matter. 
This effect provides us with a unique opportunity for testing 
gravity, using observational quantities measured through the 
gravitational lensing.

In the present paper, we focused our investigation on the 
density profile of cluster halo observed with gravitational lensing. 
Utilizing the observational data of the surface mass density 
in~\cite{Umetsu2011b} and the differential surface mass density 
in~\cite{Oguri2011}, we obtained the constraint on the parameters 
$\mu$ and $\epsilon$ for the first time, assuming
the NFW profile, the gNFW profile and the Einasto profile
as the halo density profile. We obtained the constraint, 
for instance, $-0.4\simlt\mu\simlt0.6$ for $\epsilon=0.1$, 
and $-1.5\simlt\mu\simlt2.5$ for $\epsilon=1.0$, 
respectively, at $1$-sigma confidence level, assuming the NFW profile 
(see figure \ref{fig:cont.e_m.Umetsu.NFW}).
The constraint on $\mu$ is better for small $\epsilon$, but
the constraint on $\mu$ becomes rather weak as $\epsilon$ becomes larger.
Newtonian gravity is favored when the NFW profile is assumed. 
This result is not altered when the gNFW profile is assumed 
(see section \ref{sec:results}). 
However, the modified gravity is favored than Newtonian gravity
when the Einasto profile with large $\Gamma$ is assumed
(see section \ref{sec:discussion}). 
The constraint obtained with our method is not very tight. 
The original galileon model is not excluded in our test.
But this method provides us with a unique chance to test the gravity 
theory on halo scales with cluster surveys, such as Subaru Hyper 
Suprime-Cam survey~\cite{HSC}, CLASH~\cite{CLASH}, 
LoCuSS~\cite{LoCuSS}, and XXM-XXL~\cite{XMMXXL}.

In the present paper, we assumed the well-known density profiles
of halo, which should be carefully tested with N-body simulations. 
In reference~\cite{dynamicalmass}, properties of halos in the DGP 
modified gravity model are investigated. The DGP model possesses the
Vainshtein mechanism similar to the galileon model.\footnote{
In the DGP model the relation connecting the lensing potential and
the density perturbation is not altered, therefore it is difficult
to obtain a constraint with the method in the present paper. }
From the N-body simulations in reference~\cite{dynamicalmass}, 
it is suggested that the NFW profile describes the density 
profile of halos in the DGP model though the parameters of 
the profiles are scaled. This supports our assumption.
However, in order to make the constraint robust, we need to 
check the modeling of the halo profile in future.  
As we considered the simple halo model in our analysis, our modeling 
of halo is valid only within the radius smaller than a few Mpc. 
In order to use the observational data including larger radius 
\cite{Mandelbaum06}, we need a more sophisticated formulation, 
which is also left as a future problem.

%\vspace{3mm}
%=========================================================================
%=========================================================================
\section*{Acknowledgments}%\acknowledgments
%=========================================================================
We thank Keiichi Umetsu and Masamune Oguri for providing us with 
their data and useful comments. 
We thank Rampei Kimura and Gen Nakamura for useful discussions. 
We thank Jun'ichi Yokoyama and his collaborators for many useful discussions.
We thank Claudia de Rham and Alister W. Graham for useful comments. 
We also thank Vera Richter and Chris Knobler for useful comments and supports.  
This work was supported by Japan Society for Promotion of Science 
(JSPS) Grants-in-Aid for Scientific Research (Nos.~21540270,~21244033) 
and JSPS Core-to-Core Program ``International Research Network for 
Dark Energy''. 
T.N. was supported in part by a Grant-in-Aid for JSPS Fellows.

%=========================================================================
\appendix
%=========================================================================

%=========================================================================
%=========================================================================
\section{The coefficients in the perturbation equations}
\label{sec:coefficient}
%=========================================================================
In this appendix, we present the general expressions of 
the coefficients in the perturbation equations in a 
subclass of general second-order scalar-tensor 
theory with the action (\ref{action}).
Following~\cite{GGI}, we introduce  
\begin{eqnarray}
A_0&=&{\dot{\Theta}\over H^2}+{\Theta\over H}-F-2{F_\phi\dot{\phi}\over H}-{{\cal E}+{\cal P}\over2H^2}, \\ \nonumber
A_1&=& {F_\phi\dot{\phi}\over H}, \\ \nonumber
A_2&=& F-{\Theta\over H}, \\ \nonumber
B_0&=& {\dot{\phi}^3G_X\over 2H}, 
\end{eqnarray}
where
\begin{eqnarray}
{\cal E} &=& 2XK_X -K +6HX\dot{\phi}G_X -2XG_\phi-3H^2F-3H\dot{\phi}F_\phi, \\ \nonumber
{\cal P}&=&K-2X(G_\phi+\ddot{\phi}G_X)+(3H^2+2\dot{H})F+(\ddot{\phi}+2H\dot{\phi})F_{\phi}+2XF_{\phi\phi}, \\ \nonumber
\Theta&=&-X\dot{\phi}G_X+HF+\dot{\phi}F_\phi/2, \\ \nonumber
\end{eqnarray}
and $F_\phi=dF(\phi)/d\phi$, $G_X=\partial G/\partial X$ and $G_\phi=\partial G/\partial \phi$. 
Finally, we have the following expressions for the coefficients in section 2,
\begin{eqnarray}
\alpha &=& {F_\phi\over F}\phi,
~~~~~
\xi = {2X G_X - F_\phi\over2 F}\phi,
~~~~~
\zeta = {2(A_1+A_2)H\over \beta\dot{\phi}\phi},
~~~~~
\lambda^2 = {B_0H\phi\over \beta X\dot{\phi}}, 
\nonumber \\
\beta &=& -\left(A_0+A_2{F_\phi\dot{\phi}\over FH}+(A_1+A_2){A_2\over F}\right){2H^2\over \dot{\phi}^2}. 
\label{coefff}
\end{eqnarray}

%=========================================================================
%=========================================================================
\section{Galileon model}
\label{sec:Galileon}
%=========================================================================
We consider the galileon model in curved spacetime with the action~\cite{DeffayetKGB},
\begin{eqnarray}
S=\int d^4x\sqrt{-g}\left[{{\mpl2\over 2}R} -X -{r_c^2\over M_{\rm Pl}}X\square\phi+{\cal L}_{\rm m}\right],
\end{eqnarray}
which is obtained by taking $F(\phi)=M_{\rm Pl}^2$, $K(X)=-X$ and $G(X)={(r_c^2/{M_{\rm Pl}})}X$, 
where $r_c$ is the parameter. 
This model admits a late-time de-Sitter attractor in a flat FRW 
universe.
The modified Friedmann equation on the attractor can be 
written as
\begin{eqnarray}
\left({H(a)\over H_0}\right)^2={1\over2}
\biggl[\Omega_{0}a^{-3}+\sqrt{\left(\Omega_{0}a^{-3}\right)^2+4(1-\Omega_{0})}\biggr].
\end{eqnarray}
The parameter $r_c$ is related to the cosmological parameters by 
$r_c=1/(54(1-\Omega_0))^{1/4}H_0^{-1}$.
In this galileon model,
$\alpha$, $\xi$, $\zeta$, $\lambda^2$ and $\beta$ in (\ref{coefff}) are given
by the background expansion history, as follows,
\begin{eqnarray}
&&
\alpha=0,
~~~~~
\xi = M_{\rm Pl}^{-2} X\phi G_X,
~~~~~
\zeta= {2XG_{X}\over\beta\phi},
~~~~~
\lambda^2= {\phi G_{X} \over \beta},
~~~~~
\nonumber\\
&&\beta= -1+2G_{X}(\ddot{\phi}+2H\dot{\phi})-2M_{\rm Pl}^{-2} X^2G_{X}^{2}.
\end{eqnarray}
For this galileon model along the attractor solution, it is useful to rewrite the combinations 
$\xi\zeta$ and $\lambda^2\zeta$ in terms of the matter density parameter 
$\Omega_m=\rho_m(a)/3\mpl2 H^2(a)$~\cite{KimuraKGB,Galihalo}, 
\begin{eqnarray}
\xi\zeta={(1-\Omega_m)(2-\Omega_m)\over \Omega_m(5-\Omega_m)},
~~~~~
\lambda^2\zeta=\left({2-\Omega_m\over H\Omega_m(5-\Omega_m)}\right)^2.
\label{eq:xizeta}
\end{eqnarray}
In this model, we have
$\mu=0.26$ and $\epsilon=0.53$ 
($\mu=0.19$ and $\epsilon=0.43$) 
at $z=0.32$ ($z=0.47$), which corresponds to the mean redshift of the 
clusters to measure the surface mass density by Umetsu et al. 
in~\cite{Umetsu2011a,Umetsu2011b} (the differential surface mass 
density by Oguri et al. in~\cite{Oguri2011}), respectively.

%=========================================================================
%=========================================================================

\end{document}